\documentclass[floatfix,%
 aip,
 amsmath,amssymb,
 reprint,%
 nofootinbib
]{revtex4-1}
\usepackage[utf8]{inputenc}
\usepackage[T1]{fontenc}

\usepackage{graphicx} 
\usepackage{dcolumn} 
\usepackage{bm} 
\usepackage{ragged2e}
\usepackage[compatibility=false]{caption}

\usepackage[T1]{fontenc}
\usepackage[utf8]{inputenc}
\usepackage{mathptmx}
\usepackage{etoolbox}

\usepackage{hyperref}
\hypersetup{
    colorlinks=true,
    linkcolor=blue,
    citecolor=black,
    filecolor=magenta,      
    urlcolor=cyan,
    pdftitle={Overleaf Example},
    pdfpagemode=FullScreen,
    }

\usepackage{tikz}
\usepackage{xcolor}

\makeatletter
\def\@email#1#2{%
 \endgroup
 \patchcmd{\titleblock@produce}
  {\frontmatter@RRAPformat}
  {\frontmatter@RRAPformat{\produce@RRAP{*#1\href{mailto:#2}{#2}}}\frontmatter@RRAPformat}
  {}{}
}%
\makeatother
\begin{document}

\preprint{AIP/123-QED}

\title[]{Coarse-grained simulations of dsDNA polycatenanes and network formation in annular nanochannels with topoisomerase~II}
\author{C. Palombo}
\author{D. Breoni}%
\author{L. Tubiana}
\email{luca.tubiana@unitn.it}
\affiliation{ 
Department of Physics, Universit\`a di Trento, Via Sommarive 14, I-38123 Trento, Italy 
}
\vspace{0.2em}
\affiliation{INFN-TIFPA, Trento Institute for Fundamental Physics and Applications, I-38123 Trento, Italy}

\date{\today}
             
\vspace{0.4em}

\begin{abstract}
We numerically investigate the behavior of a system of initially unlinked nicked dsDNA rings confined into an annular square nanochannel in the presence of Topo~II. Channel confinement can enhance the catenation likelihood by bringing highly bent regions from different rings in close proximity, while at the same time reducing the emergence of knots. The annular channel topology simplifies the characterization of the system by removing periodic boundary conditions and can lead to the formation of circular catenanes.
We characterize the equilibrium and dynamical properties of the steady-state system, including  the amount of catenation and the topologies explored by the system, under different parameters of the model and for different levels of confinement.  We argue that a similar setup could allow for a direct comparison between experiments and simulations under well characterized and controlled conditions, thus providing a way to select and tune computational models of Topo~II, as well as provide a way to obtain dsDNA interlocked materials in a controlled fashion. 
\end{abstract}

\maketitle


\section{Introduction}
\label{par:introduction}
Interlocked ring polymers materials such as polycatenanes~\cite{Liu2022}, 2D Olympic networks~\cite{Chen1995,Klotz2020,He2023, michieletto2025kinetoplast-937, Klotz2024,Marquez2024}, and Olympic gels~\cite{gennes1979scaling-71e,vilgis1997elasticity-10c,speed2025assembling-e4b,krajina2018active,Ubertini2021} are held together by topological links instead of covalent bonds. Numerical as well as experimental studies have shown that the additional degrees of freedom of the rings and the interplay between topological constraints and geometry provide these systems with exotic physical features, such as high deformability~\cite{Chen1995,Klotz2020,He2023,michieletto2025kinetoplast-937,orlandini2022topological}, unique viscoelastic properties~\cite{vilgis1997elasticity-10c,speed2025assembling-e4b}, and tunable flexibility~\cite{Wu2017}. Topological effects can also arise from the chirality of the links, which can affect the properties of  circular catenanes~\cite{Tubiana2022} and 2D Olympic networks~\cite{Klotz2024,Marquez2024}.  These characteristics pave the way for a wide range of potential applications, from sensors and nanomachines to candidates for novel smart materials~\cite{Xiong2021, Linko2015, Kay2007, Stoddart2017}.  

While advances in supramolecular assembly~\cite{Datta2020}, click chemistry~\cite{Bunha2015click} and metallo-template synthesis~\cite{Wu2017} have made it possible to build linear and branched polycatenanes of up to hundreds of small rings, it is still challenging to obtain large Olympic gels and networks, as well as polycatenanes with controlled overall topology (e.g. specific branching, or circular ones). A promising avenue in this direction leverages the properties of dsDNA rings either through the enzymatically controlled cyclization of thousand of different dsDNA molecules~\cite{speed2025assembling-e4b} or by using type~2 topoisomerases (Topo~II) enzymes to actively catenate and decatenate dsDNA plasmids, leading to a material whose properties can be controlled by changing the enzyme activity, as demonstrated by Krajina et al~\cite{krajina2018active}. The latter approach is inspired by the kinetoplast DNA (kDNA), a network of thousands of interlocked dsDNA rings that form the mitochondrial DNA of Trypanosomatids~\cite{Chen1995,Klotz2020,He2023,ramakrishnan2025organisation,Diggines2024,michieletto2025kinetoplast-937} and whose catenation is controlled by the action of topoisomerase enzymes, with such a precision that, on average, each ring is linked to three others~\cite{Chen1995,He2023}. 

Topo~II enzymes play a central role in regulating DNA topology in living systems in general. They mediate the relaxation of torsional stress, the resolution of complex topological states and the maintenance of large-scale genome organization~\cite{Sikorav1994,Champoux2001,christensen2002dynamics-221,samejima2012mitotic-cf1,vologodskii2016disentangling-c92}. While  Topo~II is known to unknot and decatenate dsDNA, experimental results have shown that in the presence of polyvalent counterions~\cite{krasnow1982catenation-349}, specific other proteins~\cite{riou1985type-ii-6d0}, or simply of a high concentration of Topo~II~\cite{Jeong2022}, the enzyme's behavior switches to knotting and catenating the rings. In the latter case, this has been linked to the presence of an intrinsically disordered C-terminal domain in in several types of Topo~II (ScTopo~II, Topo~II$\alpha$, Topo~II$\beta$) that induces a condensation stimulated by the presence of DNA~\cite{Jeong2022}. Note that Topo~II$\alpha$ is the enzyme used by Krajina et al.~\cite{krajina2018active} to obtain an Olympic gel whose rheological properties could be controlled by regulating the enzymatic activity. 

Several computational studies have proposed coarse-grained models of Topo~II activity, either for decatenation/unknotting~\cite{flammini2004simulations-b7c,rybenkov1997simplification-a83,vologodskii2016disentangling-c92,liu2006topological-d0b,witz2011tightening-1a1,Rawdon2016,Michieletto2022} or, more recently, to model a setup such as that of Krajina et al.~\cite{krajina2018active}, where Topo~II both catenates and decatenates dsDNA~\cite{Ubertini2021,zhang2024olympic-594,zhang2025dna-a3d}. Most models treat Topo~II implicitly~\cite{Ubertini2021,Michieletto2022,zhang2024olympic-594}, for example by randomly selecting some suitable portion of dsDNA and making them amenable to strand passage, either by enlarging a bond~\cite{zhang2024olympic-594}, or by making them soft~\cite{Michieletto2022,battaglia2025binding-ca6} for a limited amount of time. The selection of the strands can be biased to account for known and supposed preferences of Topo~II, such as highly bent dsDNA regions~\cite{witz2011tightening-1a1}, or hook-like juxtapositions~\cite{liu2006topological-d0b,liu2015consistent-f8f,vologodskii2016disentangling-c92}. Unfortunately, due the complexity of Topo~II-related experiments, as well as the limits inherent in coarse-grained computational models, it is difficult to systematically test the realiability of these models in predicting the formation of dsDNA interlocked materials.

In this manuscript we simulate the formation of dsDNA polycatenanes and networks through Topo~II activity when the system is subject to confinement into an annular nanochannel. Such nanochannels represent a novel experimental setup which has been used, for example, to study self-ligation of long DNA molecules~\cite{Berard2016} and that, we believe, could provide both a setup to test Topo~II models and for the controlled production dsDNA polycatenanes.

Nanochannel confinement makes it possible to leverage several experimental and computational results on relation between topological and geometrical properties of confined polymer rings~\cite{Micheletti2012,Micheletti2014,Suma2015,jain2017simulations-2df,ma2020diffusion-980,ma2021diffusion-824,tubiana2024topology-7cd}. The annular topology effectively removes boundary conditions, simplifying the analyses, and has the added benefit of allowing the formation of circular polycatenanes and networks encompassing the whole channel. 
Given the estimated diameter of Topo~II cores of $\sim15$nm~\cite{schultz1996structure-d15}, we assume that channels with a square cross section of side $\Delta_r\geq 25~\rm nm$, compatible with the onset of the Odijk regime for dsDNA rings~\cite{benkova2012simulation}, will be large enough to host Topo~II enzymes. At the same time, these narrow channels considerably hinder the formation of knots~\cite{Micheletti2012} and keep dsDNA plasmid oriented along the channel. By controlling the amount of dsDNA and Topo~II it is possible to keep the dsDNA rings close to each other and to bring the highly bent regions of different plasmids to overlap, thus, in principle, favoring the catenating action of the enzyme in a controlled fashion. 

In order to provide a set of baseline results showing what can be expected in the setup described above, we use Langevin dynamics and a coarse-grained Topo~II model to simulate the evolution of a system of initially unlinked dsDNA rings confined inside an annular channel. Our model mixes elements of two recent ones, that of Michieletto et al.~\cite{Michieletto2022} and the one by Zhang et al.~\cite{zhang2024olympic-594,zhang2025dna-a3d}, and has been tuned to avoid multiple-strand passages. 
We investigate the structural and dynamical properties of the system for different model parameters and channel widths, in particular the amount of  catenated regions, their span and dynamics, and we characterize the most common topologies, identifying the conditions that can lead to the formation of circular polycatenanes, spanning the whole channel, exploiting the annular topology of the channel. 


The manuscript is organized as follows: Section~II describes the model implementation and the set of observables used to characterize the system, Section~III presents the results and their interpretation, and Section~IV summarizes our conclusions.

\section{Methods}
\label{par:methods}
\subsection{Simulated system}
We consider systems of $M$ $\sim$1 kbp long nicked dsDNA rings, modeled as Kremer–Grest bead-spring polymers~\cite{Kremer1990,Grest1986} composed of $N=130$ beads. The diameter of the beads, $\sigma$,  is assumed to be $\sigma = 2.5~\rm nm$ (giving $955~\mathrm{bp}$ per ring), consistent with the effective thickness of double-stranded DNA in solution at high salt concentrations~\cite{rybenkov1993probability}. Chain connectivity is treated with the finitely extensible nonlinear elastic (FENE) potential
\begin{equation}
    U_{\mathrm{FENE}}(r) = -\frac{k}{2} R_0^2 \ln \left[ 1 - \left( \frac{r}{R_0} \right)^2 \right] \,,
    \label{eq:FENE}
\end{equation}
where $r = |\mathbf{r}_{i+1}-\mathbf{r}_i|$ is the distance between two consecutive beads having coordinates $\mathbf{r}_i$ and $\mathbf{r}_{i+1}$, $k = 30\epsilon/\sigma^2$ and $R_0 = 1.5\sigma$. 
The stiffness of dsDNA is modeled via a Kratky–Porod bending potential~\cite{KratkyPorod1949}
\begin{equation}
    U_{\rm bend} = k_{\rm bend} (1-\cos \theta) \,,
    \label{eq:bending}
\end{equation}
where $\theta$ is the angle defined by two consecutive bond vectors along the chain contour. The bending stiffness is set to $k_{\rm bend} = 20\,\frac{\epsilon}{\sigma}l_p$. This choice corresponds to a persistence length $l_p = 50~\rm nm$~\cite{Bustamante1994}.
Inter-bead interactions are modeled with a Weeks-Chandler-Andersen (\text{WCA}) potential~\cite{Weeks1971}:
\begin{equation}
        U_{\mathrm{WCA}}(r) = 4 \epsilon \left[ \left( \frac{\sigma}{r} \right)^{12} - \left( \frac{\sigma}{r} \right)^{6} + \frac{1}{4}\right] \theta \left(r_c - r \right)
    \label{eq:WCA} \,,
\end{equation}
where $r$ is the distance between any pair of beads and the cutoff is $r_c = 2^{1/6}\sigma$.\\

Simulations are performed using the \textsc{LAMMPS} molecular dynamics engine~\cite{plimpton1995fast}, employing Langevin dynamics in reduced Lennard–Jones units, where $\sigma$ is the bead diameter, $\epsilon = k_BT$ sets the energy scale, and $m$ is the monomer mass. The integration time is set to $\text{d}t = 0.0124\tau_{LJ}$ and the Langevin damping time to $\tau_{\rm damp}=\tau_{LJ}$. 
 
\begin{figure}[h!]
    \centering
    \includegraphics[width=0.95\columnwidth]{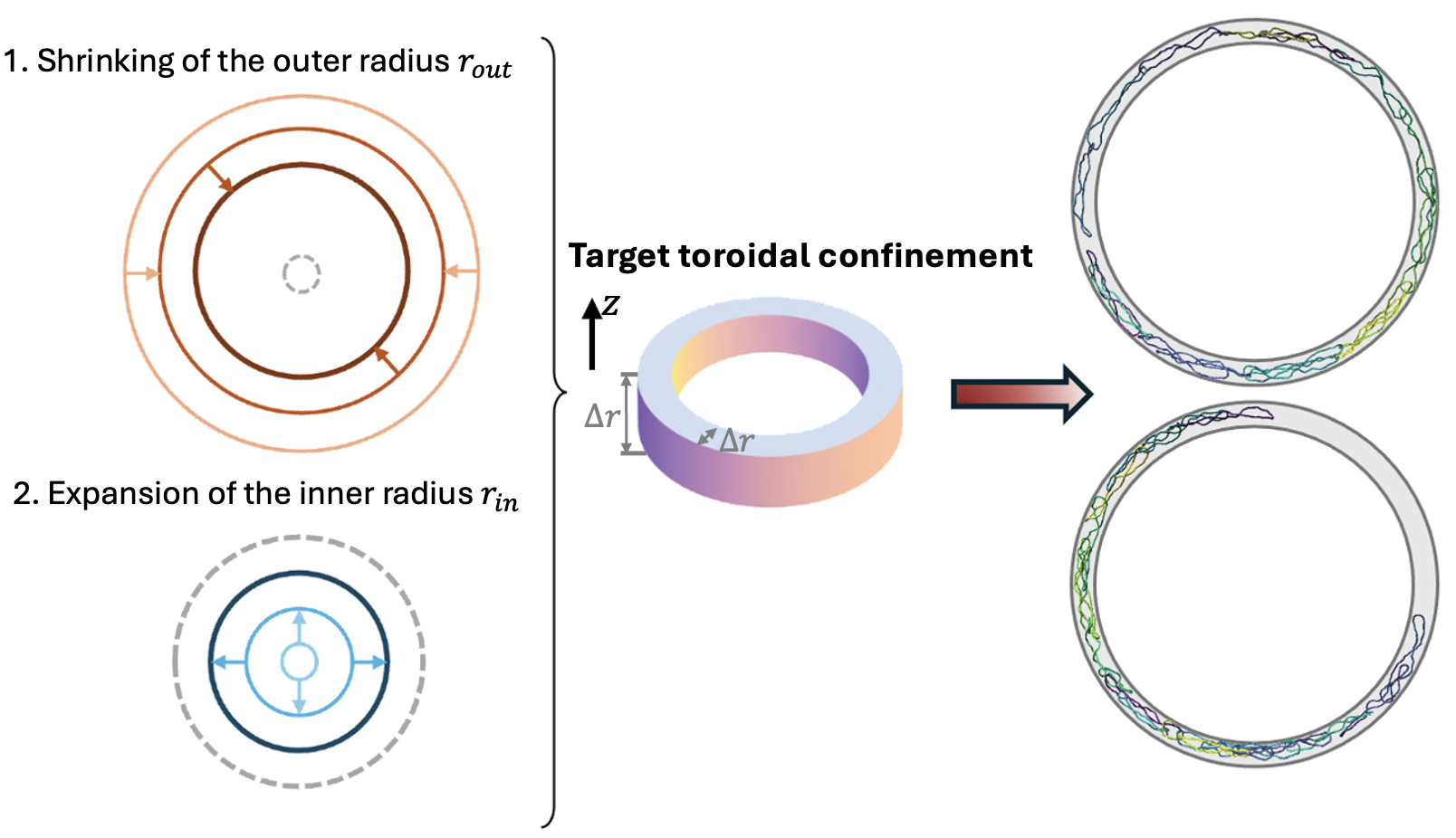}
    \caption{Schematic visualization of the initialization procedure of the confinement, together with two different polymer configurations. Left: wall shrinking procedure used to obtain the initial configurations. The outer wall is first slowly shrunk to its desired final radius, followed by the expansion of the inner wall radius. The height of the channel is kept fixed. Center: the confining channel. Right: two example initial configurations for a system with $M=15$ rings, in a channel of $\Delta r = 10\sigma$ and $r_m=85\sigma$ (top-down view).}
    \label{fig:confinement}
\end{figure}

The annular channel is bounded in the $xy$ plane between two concentric cylinders of radii $r_{\rm in}$ and $r_{\rm out}$, whose axis lie along $z$, such that the thickness of the channel is $ r_{\rm out} - r_{\rm in}\equiv \Delta r$. In $z$, the channel is limited by two planar walls placed on the $xy$ plane at heights $\pm \Delta r/2$, yielding a square cross section (see Fig.~\ref{fig:confinement}, center).  
We consider in particular two thickness sizes: $\Delta_r = 20\sigma$ ($50~\rm nm$) and $\Delta r = 10\sigma$ ($25~\rm nm$). 
A range of confinement volumes is explored by varying $r_m = \frac{1}{2}(r_{\rm in} + r_{\rm out})$. 
Confinement is imposed through a standard Lennard-Jones potential between the walls and the polymers' beads, using LAMMPS' \href{https://docs.lammps.org/fix_wall.html#fix-wall-lj126-command}{\texttt{fix wall/lj126}} command. 

To produce unbiased initial configurations of $M$ rings, we start from a low confinement condition with the height of the channel set to its desired value,  but with the cylinders set aside by $r_{\rm out} -r_{\rm in} = 50\sigma$ -- slightly larger than the diameter of the rings in their initial circular configuration --  and gradually brought closer until their distance reaches the desired value of $\Delta_r$ (Fig.~\ref{fig:confinement}, left).


\subsection{Topo~II}
\begin{figure}[h!]
    \centering
     \includegraphics[width=0.98\columnwidth]{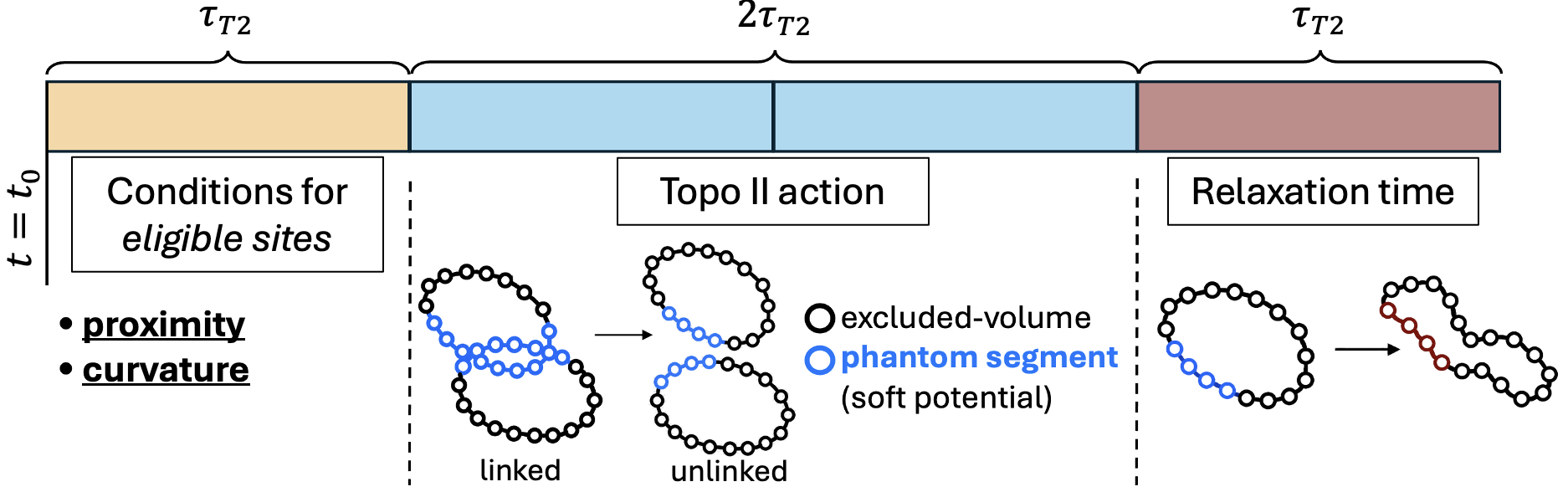}
    \caption{Scheme representing the implicit computational model of Topo~II.}
    \label{fig:topo_scheme}
\end{figure}

Our model for Topo~II takes inspiration both from the model of Michieletto et al.~\cite{Michieletto2022} and that of Zhang et al.~\cite{zhang2024olympic-594,zhang2025dna-a3d}. From the first, we take the fact that the action of Topo~II is mimicked by strands becoming soft; similar to the latter, we model Topo~II activity with a three step process corresponding to the identification of possible sites, the activation of a chosen site, and a relaxation time in which the same site cannot be chosen, see Fig.~\ref{fig:topo_scheme}. The duration of these steps is regulated by a characteristic time $\tau_{\rm T2}$, with the identification and relaxation step lasting $\tau_{\rm T2}$ and the action step lasting $2\tau_{\rm T2}$. We performed simulations with either $\tau_{\rm T2}=1240 \tau_{LJ}$ or $\tau_{\rm T2}=100 \tau_{LJ}$.
The Topo~II activity loop is implemented by calling LAMMPS from Python. Every $\tau_{\rm T2}/\text{d}t$ steps the Python script modifies the LAMMPS parameters to define the new active sites and those that enter the relaxation period, restarting then the simulation.
\label{par:topo_model}

In the first Topo~II activation step, that is the identification of sites eligible for crossing, sites are chosen based on the proximity of two dsDNA strands or on proximity together with local curvature. Sites eligible for strand passage events are identified when two beads remain within a cutoff distance of
\[
r_{\rm thr} = 2.0\sigma
\]
for at least $\tau_{T2}/dt$ steps. When present, the condition on curvature further adds the requirement that the strands containing the two proximal beads have a local curvature above a certain threshold. We compute the local curvature around a bead $\mathbf{r}_i$ as:
\begin{equation}
    \kappa_i^{(m)} = \frac{\lVert \mathbf{r}_{i+m} - 2\mathbf{r}_i + \mathbf{r}_{i-m} \rVert}{\lVert \mathbf{r}_{i+m} - \mathbf{r}_{i-m} \rVert^2} \,,
    \label{eq:curvature}
\end{equation}
where $m=4$, so that $\kappa_i^{(m)}$ gives the curvature of a strand having $2m+1=9\sigma$ ($22.5~\rm nm$), with bead $i$ at its center. This size was chosen of the same order of experimentally measured bent regions~\cite{Peng1995}. We choose a threshold $\kappa_{thr} = 0.08~\sigma^{-1}$, corresponding to an effective bending angle of 
approximately $58.3^\circ$. 

At the end of this first step, Topo~II activation events are generated by extracting the number of possible events $n_{\rm T2}$ from a Poisson distribution whose mean is proportional to the total number of beads, yielding a constant per-bead activity. The distribution mean is chosen such that, on average, one event per $\tau_{\rm T2}$ is attempted every $400$ beads. For simplicity, we consider Topo~II enzymes to be in a high enough concentration as to be always available. 

When the proximity criterion alone is considered, $n_{\rm T2}$ sites are selected randomly from the eligible ones. Conversely, when also the additional curvature condition is considered, we select the $n_{\rm T2}$ from a probability distribution in which each site is assigned a weight proportional to the local curvature of the two strands:
\begin{equation}
\label{eq:weight_curvature}
w_{ij} = 1 + \alpha [(\kappa_i -\kappa_0)_+ + (\kappa_j - \kappa_0)_+]\,,
\end{equation} 
where $(x)_+ \equiv \max(0,x)$, $\kappa_0$ is the curvature threshold and $\alpha = 2$.\\

The selected Topo~II sites are activated in the second step. The selected strands continue to interact with other portions of the rings through the WCA potential of eq.~\ref{eq:WCA}, but interact with one another through the soft potential~\cite{Michieletto2022}:
\begin{equation}
    U_{\text{soft}} =
    \begin{cases}
        A \left( 1 + \cos\left( \dfrac{\pi r}{r_c} \right) \right) & r < r_c = 2^{1/6}\sigma \\
        0 & \text{otherwise}.
    \end{cases}
    \label{eq:soft}
\end{equation}
For each triggered event, the soft potential is applied to two local segments extending ($\pm 6$) beads from the contact beads, resulting in a $13$-bead soft window on each strand. 
Preserving the excluded-volume interactions between the selected strands and the non-activated portions of the rings considerably reduces the possibility of multiple-strand passages, which are incompatible with the biological action of Topo~II. 
The soft interaction is implemented with a mild repulsion ($A = 2$), which is reinforced ($A = 20$) for the $5 \times 10^4$ timesteps just before the reactivation of the WCA potential. This is done to ensure numerical stability and avoid unphysical bead overlaps upon restoring excluded-volume interactions. 

Finally, the third step implements a relaxation period that  prevents the reactivation of the same sites.

\subsection{Observables and topological analyses}\label{par:top_analyses}
All simulations with $\tau_{\rm T2} = 1240 \tau_{LJ}$ are run for a time $1.2\times10^7\tau_{LJ}$, ensuring that during the simulation at least $10^4$ activation windows are explored, in order to sample a large number of different topologies. For $\tau_{\rm T2} = 100\tau_{LJ}$, due to the more frequent communication between Python and LAMMPS, the simulations are run up to $0.6\times10^7,\tau_{LJ}$, which corresponds to approximately $6\times 10^4$ activation windows. Geometrical and topological properties are evaluated on configurations dumped every $7\times 10^4$ timesteps ($868\tau_{LJ}$) for $\tau_{\rm T2} = 1240 \tau_{LJ}$ and $1.9\times 10^4$ timesteps ($234\tau_{LJ}$) for $\tau_{\rm T2}=100\tau_{LJ}$, in order to map the dynamics as well. The frequency is limited by storage considerations.

The topological status of catenated ring configurations is identified by computing the Jones Polynomial with Topoly~\cite{topoly}. Since our channels are quite narrow,  the analysis is restricted to pairwise links only, without considering higher-order structures. 
From each configuration, the \emph{unlinking number} has then been obtained through tabulated values available in the work of Kohn~\cite{unlink}.
From the list of linked rings, we constructed networks using the Python package networkX~\cite{hagberg2020networkx} version 3.4.2. We characterized the connected components of such networks  through two topological indices, the Wiener index $W$ and the cyclomatic number. 
Considering a graph $G(V,E)$ having $V$ vertices and $E$ edges, the Wiener index $W(G)$ is defined as 
\begin{equation}
    W(G) = \sum_{i<j}d_{ij},
\end{equation}
where the sum runs over all vertices of the graph and $d_{ij}$ is the length of the shortest path connecting vertices $i$ and $j$ ($d_{ij}=\infty$ if no connection is present). If the graph $G$ is connected, the Wiener index lies between the minimum value
\begin{equation}
    W_{min}(G) = \frac{V(V-1)}{2},
\end{equation}
obtained when $G$ is complete (every vertex has distance one from every other vertex) and the maximum
\begin{equation}
    W_{max}(G) = \frac{V(V^2-1)}{6},
\end{equation}
obtained when $G$ is a linear chain. 

The cyclomatic number $n_{loops}$ counts the number of independent loops present in $G(V,E)$. Considering again a connected graph, $n_{loops}$ is equal to:
\begin{equation}
    n_{loops} = E-V+1,
\end{equation}
so that its minimum is zero for trees and linear graphs and its maximum is 
\begin{equation}
    n_{loops}^{max} = \frac{(V-1)(V-2)}{2},
\end{equation}
if $G(V,E)$ is complete.

\section{Results}
\begin{figure}
    \centering
    \includegraphics[width=0.98\linewidth]{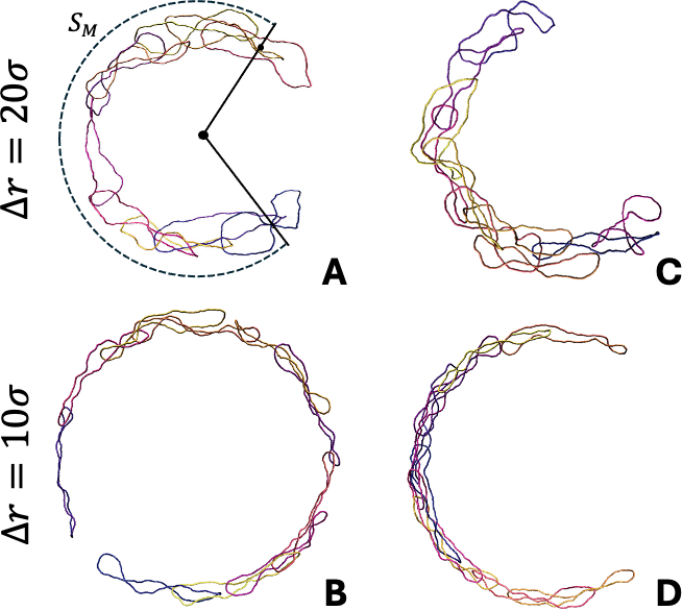}
    \caption{Rendering of four different simulation snapshots with $M=10$ for $\Delta r = 20\sigma$ (A, C) and $\Delta r = 10\sigma$ (B, D), comparing well-distributed states (A, B) and accumulated ones (C, D). Panel A also graphically shows the definition of the \textit{covered angle} $S_M$.}
    \label{fig:S_M rendering}
\end{figure}
As described in Sec.~\ref{par:methods}, we simulate systems of $M$ rings in annular (toroidal) square channels having mean radius $r_m=\frac{1}{2}(r_{out}+r_{in})$ and two different widths: $\Delta_r = 20\sigma$ ($50~\rm nm$) and $\Delta_r=10\sigma$ ($25~\rm nm$), see Fig.~\ref{fig:S_M rendering}.  In order to compare the behavior in channels of different widths and radii, we fix the linear density of the rings or, equivalently, the angular fraction of the circumference, $f$, covered by the rings. We set $f=1.6$, so that the density of rings in the channel is always above the overlap one. This choice of $f$ allows in principle for the formation of  circular catenanes encompassing the whole channel, but is low enough to discourage the formation of highly complex structures of the same extension. Defining the $r_{\parallel,\Delta_r}$ as half the average extension of a single ring along a channel of size $\Delta_r$, the number of rings is: 
\begin{equation}
    M_{r_m,\Delta_r}=\frac{f \pi r_m}{r_{\parallel,\Delta_r}}\,.
\end{equation}
In this way, if the rings were placed on the vertices of a regular $M-$gon, they would have a given overlap $f$. 

As described in sec.~\ref{par:topo_model}, we simulate Topo~II strand-passage action through a cycle of viable site identification, activation, and relaxation. Site identification always follows a proximity condition in which two strands must be no more than $2\sigma$ apart from one another, plus an optional condition on local curvature, which is applied to a subset of replicas. The duration of these steps is characterized by the time $\tau_{\rm T2}$, as defined in sec.~\ref{par:topo_model}.  $\tau_{\rm T2}$ thus controls the activity of Topo~II and should be compared to the typical time needed by the rings to diffuse along the channel. Specifically, we consider $\tau_{\rm T2}= 1240\tau_{LJ}$, and $\tau_{\rm T2}=100\tau_{LJ}$. In the first case, the characteristic time of Topo~II is slightly longer than the time needed by a ring to diffuse more than $2\sigma$ -- the characteristic distance for Topo~II activity in our model -- while in the latter it is considerably shorter.
For each value of $r_m$, $\Delta_r$, and $\tau_{\rm T2}$ we simulated several replicas with the proximity-only activation condition and one with the proximity-plus-curvature condition.

To facilitate the interpretation of our results, we consider two reference systems. The first is  given by fully self avoiding rings, which are assumed to remain on average on the vertices of a regular $M$-gon to minimize their topological repulsion. The second is given by fully phantom rings. Having no interactions with each other, their center of masses can be assumed to be randomly distributed on the circle with uniform probability, so that other quantities like the fraction of the channel occupied by the rings can be obtained analytically.

\subsection{Angular distribution of the rings}\label{sec:ang_dist}
We first characterize the spatial distribution of rings inside the toroidal channel by measuring a quantity akin to the span in linear channels: the minimum angle containing the centers of mass of all the $M$ rings, which we call \emph{covered angle}, $S_M$ (see Fig.~\ref{fig:S_M rendering}A). This quantity ranges from $S_M=0$ when all centers overlap to $S_M = 2\pi(1-\frac{1}{M})$ when the $M$ rings are equally spaced along the channel, forming the vertices of a regular $M-$gon. The value of $S_M$ for our reference cases can be obtained analytically. For self-avoiding rings we can assume that it corresponds to its maximum. For phantom rings, the smallest angle containing $M$ points uniformly distributed on a circle has been analytically derived and is given by~\cite{Stevens1939, Huffer1987}:
\begin{equation}
\label{eq:phantom}
\left\langle \frac{S_M}{2\pi}\right\rangle = 1-\frac{H_M}{M},
\end{equation}
where $H_M=\sum_{k=1}^M \frac{1}{k}$ is the $M$-th harmonic number (see SM). 
\begin{figure}[h!]
    \centering
    \includegraphics[width=0.98\linewidth]{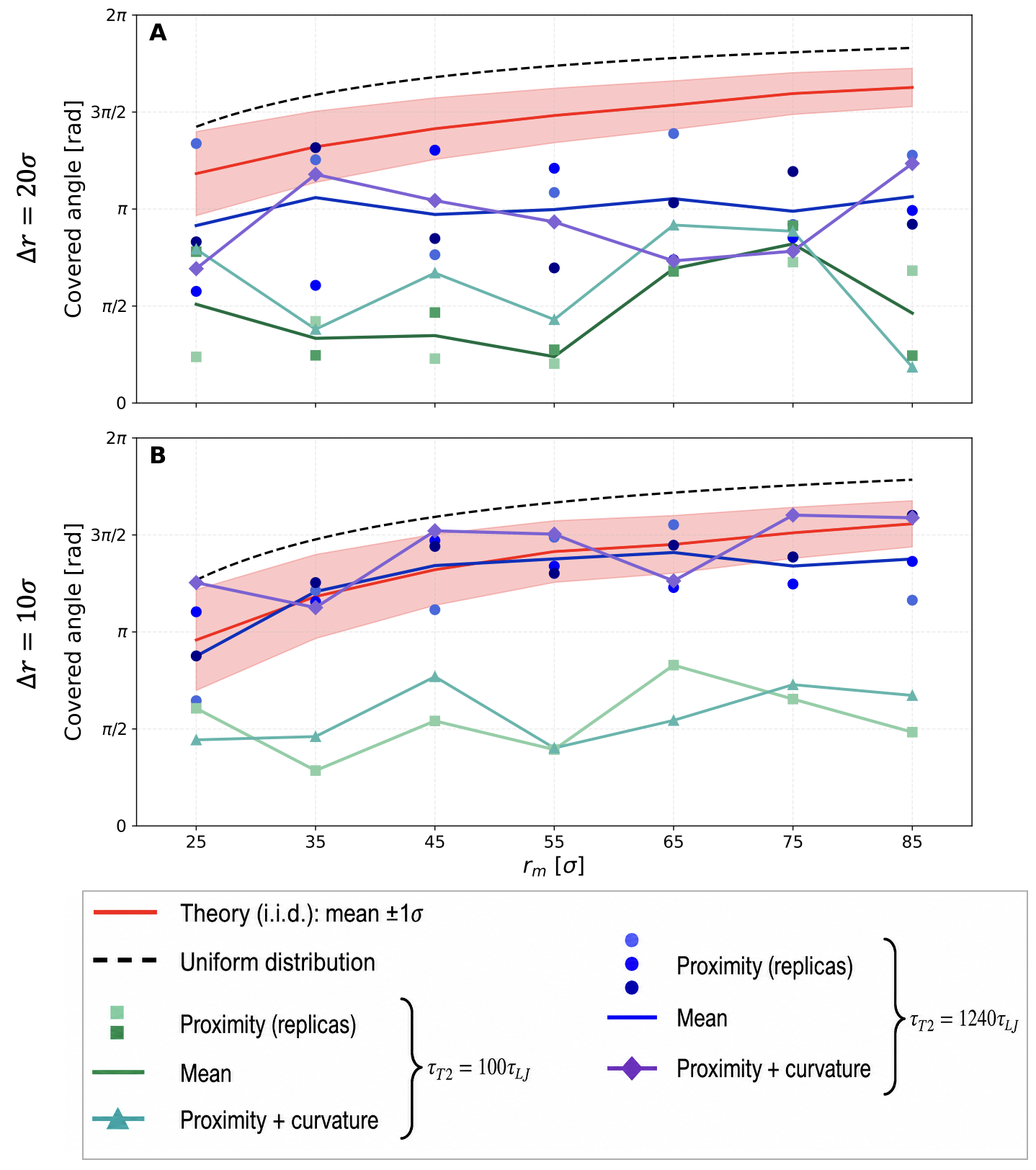}
    \caption{Covered angle, $S_M$, as a function of mean radius $r_m$. Red band: phantom theoretical prediction. Black dashed line: expected angle for self-avoiding rings. Blue (violet) markers: Topo~II simulations with proximity  (plus curvature) condition and $\tau_{\rm T2} = 1240\tau_{LJ}$. Green (aqua) markers: Topo~II simulations with proximity (plus curvature) condition and $\tau_{\rm T2} = 100\tau_{LJ}$.} 
    \label{fig:covered_angle} 
\end{figure}

As shown in Fig.~\ref{fig:covered_angle}, the total angular span of the rings, and thus their accumulation, depends on the channel width, $\Delta_r$, and on the characteristic time $\tau_{\rm T2}$ of activation of the model Topo~II. For channels of width $\Delta_r = 20\sigma$ (50 nm) the rings remain consistently more accumulated than phantom rings, and the average value of $S_M$ across replicas does not seem to increase with the number of the rings. The accumulation gets further accentuated by reducing the action time of Topo~II (see Fig.~\ref{fig:covered_angle}a, blue and green curves respectively). 

Reducing the confinement to $\Delta_r=10\sigma$ (25 nm) instead brings a qualitative distinction between the models with different values of the Topo~II action time. While the model with the faster action time, $\tau_{\rm T2}=100\tau_{LJ}$, still shows a significant accumulation, the model with longer action time, $\tau_{\rm T2}=1240\tau_{LJ}$, now follows the predictions for phantom rings, see  Fig.~\ref{fig:covered_angle}b). In both cases, including the further conditions on local curvature does not change the results significantly. 

It is interesting to look at how the spatial organization of the rings, as described by the angle $S_M$, evolves in time. Taking advantage of the circular confinement, we expand the angular density distribution function of the center of masses, $P(\theta,t)$, as a Fourier series:
\begin{equation}
\label{eq:p_theta}
P(\theta,t)=\frac{1}{2\pi}\sum_{m=-\infty}^{+\infty} c_m(t)e^{-im\theta}, \text{with }
c_m(t)=\int_0^{2\pi}P(\theta,t)e^{im\theta}d\theta.
\end{equation}

The amplitude of the first mode is
\begin{equation}
\label{eq:R1}
R_1(t)=|c_1(t)|=\left|\frac{1}{N}\sum_{j=1}^{N}e^{i\theta_j(t)}\right|\,,
\end{equation}
where $N$ is the total number of polymer beads in the system and $\theta_j(t)$ is the angular position of bead $j$ at the time $t$. 
$R_1$ is invariant under global rotations and 
can act as an order parameter for angular accumulation. If all rings are located equally far, $R_1\simeq 0$; if their center of masses are all very close, $R_1\simeq 1$. \\

In order to have a common reference for system with a different number of rings, we leverage again the equivalence between the center of masses of phantom rings and points uniformly distributed on a circle. This allows us to numerically compute $\langle R_1^{\textrm{phantom}}\rangle$ and its standard deviation $\sigma_{R_1^\textrm{phantom}}$ for a system of $M$ random points on the circle, and use them to normalize $R_1(t)$: 
\begin{equation}
    \tilde{R}_1(t) = \frac{R_1(t)-\langle R_1^{\textrm{phantom}}\rangle}{\sigma_{R_1^\textrm{phantom}}}\,.
\end{equation}
\noindent 
Using this definition, $\tilde{R}_1(t)>1$ indicates that the system is accumulated beyond what can be expected for phantom rings, while $\tilde{R}_1<-1$ indicates that the system is significantly less aggregated than phantom rings. 

\begin{figure}[ht]
    \centering
    \includegraphics[width=0.95\columnwidth]{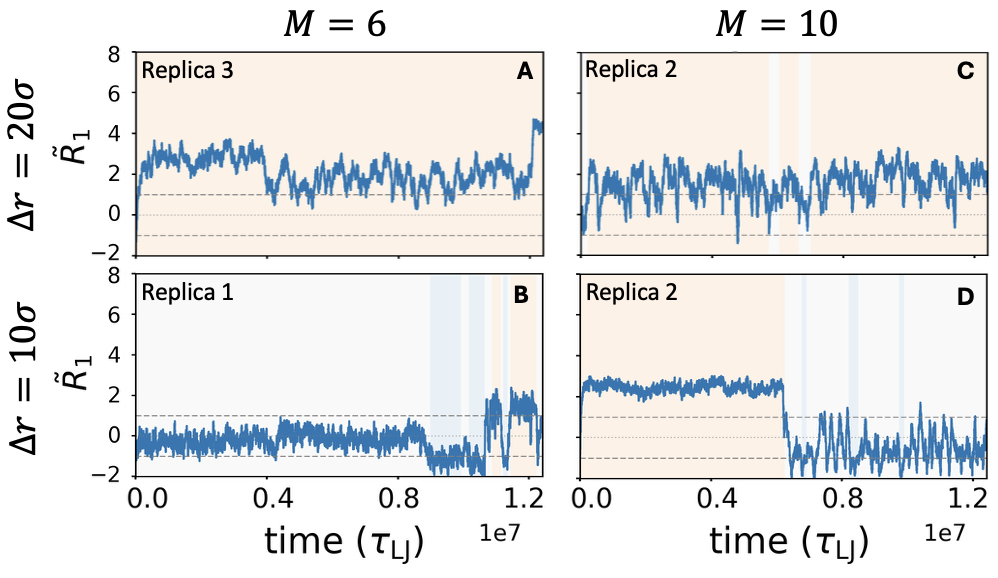}
    \caption{
    Time evolution of the normalized first Fourier mode $\tilde{R}_1(t)$ of the centers of mass of the rings for example replicas with $M=6$ and $M=10$ rings and channel width $\Delta_r=20\sigma$ (top), $\Delta_r=10\sigma$ (bottom). Gray dotted lines represent the expected value for phantom rings, gray dashed lines identify the interval $\pm \sigma$. Positive values  of $\tilde{R}_1$ indicate aggregation beyond that expected for phantom rings, while negative values indicate that the rings are well distributed along the channel. These situations are emphasized by the background color. }
    \label{fig:bimodality_Fourier} 
\end{figure}

In Fig.~\ref{fig:bimodality_Fourier}, we report the evolution of $\tilde{R}_1(t)$ for a few representative replicas for $\tau_{\rm T2}=1240\tau_{LJ}$, different system sizes and channel sizes. 
As expected from the behavior of the covering angle $S_M$ (see Fig.~\ref{fig:covered_angle}), the rings in a larger channel are more accumulated than those in a narrow one. Nonetheless, while the phenomenon is more frequent for $\Delta_r=10\sigma$, both channels show jumps from aggregated states to more free configurations in which the rings behave either as phantom chains or as more isolated chains, spanning larger portions of the channel, and vice-versa.  
Data for all other replicas, including those with the additional local-curvature condition, is reported in the SM and display a similar qualitative behavior. 

\subsection{Topological complexity}
\begin{figure}[h!]
    \centering
    \includegraphics[width=0.95\columnwidth]{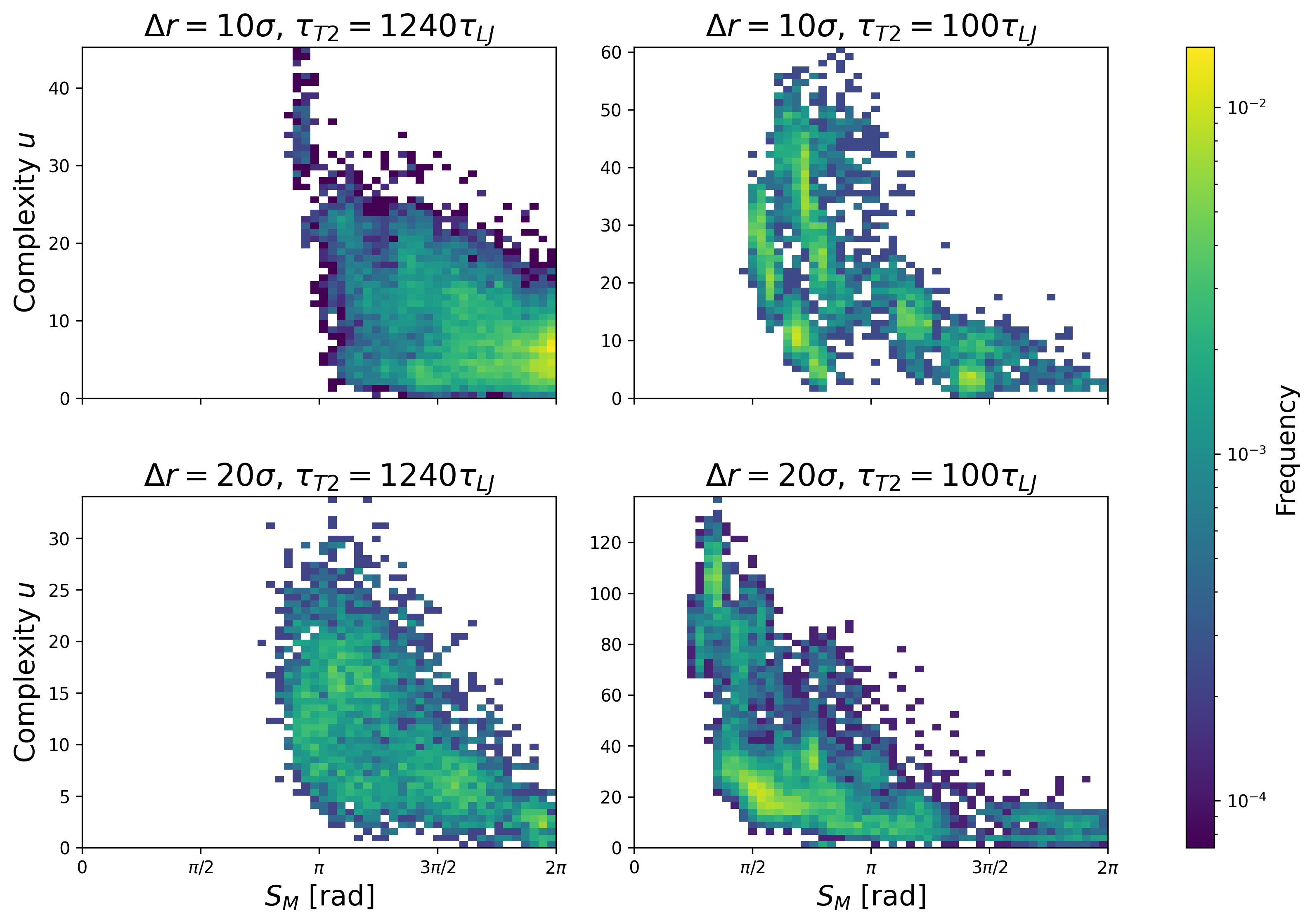}
    \caption{Unlinking number vs $S_M$ for different value of the channel width, $\Delta_r=10\sigma$ and $\Delta_r=20\sigma$, and for different times $\tau_{\rm T2}$. The heatmaps represent the frequencies of values of $u$ and $S_M$ observed during the dynamics over all proximity replicas for all values of $r_m$.}
    \label{fig:complexity}
\end{figure}
To gain insight into the mechanisms underlying the accumulation observed with larger channels and smaller Topo~II action times,  we investigate how topological complexity correlates with spatial organization. We quantify the topological complexity of a system with the \textit{unlinking number}, $u$, which measures how many strand passages are required to fully disentangle a linked configuration: the higher the linking number, the more disentangling actions from Topo~II are necessary to free the rings, see Sec.~\ref{par:top_analyses}. To understand its relation with $S_M$, we computed the 2D histogram of the frequency of $(S_M,u)$ across all values of $r_m$.
As shown in Fig.~\ref{fig:complexity} for the proximity-based Topo~II model, we observe a clear anticorrelation between $u$ and $S_M$, with higher topological complexity associated with smaller values of $S_M$.  In addition to this, narrow channels favor lower values of $u$ and higher values of $S_M$, while larger channels present more uniformly spread distributions, with configurations having $u>10$ and $S_M<\frac{3\pi}{2}$ being overall more represented than for $\Delta_r=10\sigma$. 
Comparing the distributions in the two channels for $\tau_{\rm T2}=1240\tau_{LJ}$ and $\tau_{\rm T2}=100\tau_{LJ}$ one can notice that the faster activation time leads to more accumulated systems, as seen in Fig.~\ref{fig:covered_angle}, thanks to an increase in topological complexity. This can be understood as an interplay of the Topo~II action time and the characteristic time required for the diffusion of the rings along the channel. For $\tau_{\rm T2}=1240\tau_{LJ}$ the rings are able to diffuse by about $5\sigma$ during the time the model Topo~II remains in the active state (soft strands). This ensures that activation events will have a high probability of strand passage, resulting in the rings behaving similarly to the phantom reference case. For the shorter $\tau_{\rm T2}$ instead, strand passages will effectively take place mostly when the rings are already accumulated, as the available Topo~II sites are frequently switched. Once the rings are interpenetrated, it thus becomes easier for them to remain so. Adding a curvature-based criterion to the Topo~II model does not change this quantitative behaviors, see SM.

\subsection{Catenated network topologies}
\begin{figure*}
     \centering
    \includegraphics[width=0.95\textwidth]{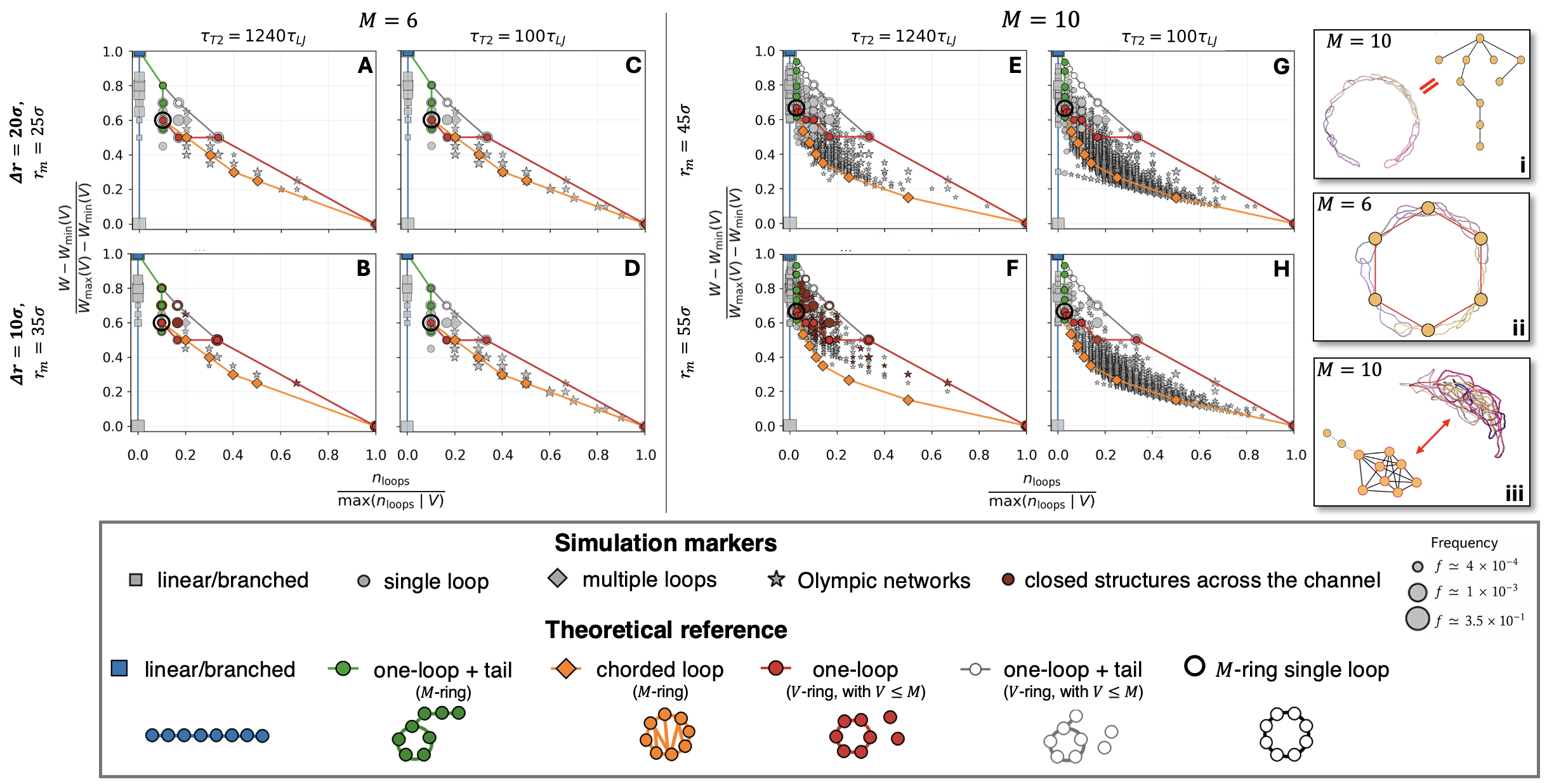}
    \caption{Different catenation networks observed over all proximity-based replicas in systems with $M=6$ (left panels: A, B, C, D) and $M=10$ (right panels:E, G, F, H) for different confinement volumes: $\Delta_r =20\sigma$ (A, C, E, G) and $\Delta_r=10\sigma$ (B, D, F, H), for $\tau_{\rm T2} = 1240\tau_{LJ}$ (A, B, E, F) and $\tau_{\rm T2} = 100\tau_{LJ}$ (C, D, G, H). Panels $\rm i-iii$ on the right show a few example network structures. Each connected component in the network is represented in panels A-H in terms of its number of independent graph loops, $n_{\rm loops}$, and its Wiener index $W$. Marker shapes distinguish linear/tree structures, single-loop structures, configurations with multiple independent loops, and Olympic networks. Marker size is proportional to the observed frequency. Dark-red symbols identify architectures that are formed across the toroidal channel (panel $\rm ii$ on the right) rather than being localized within a sector (panels $\rm i-iii$). Colored lines show the theoretical baseline of reference architectures; most configurations lie within the region bounded by these reference structures. 
    }
    \label{fig:linked_structures}
\end{figure*}
Finally, 
we analyze the catenated networks that emerge from the action of Topo~II. As described in sec.~\ref{par:top_analyses}, we limit ourselves to pairwise linking,
and map the catenated components to graphs. 
Taking inspiration from a recent approach to characterize branched polymers~\cite{vaupotivc2025normalized} we map these graphs to a two-dimensional plane whose coordinate correspond to two complementary topological indices, the Wiener index $W$ and the cyclomatic number $n_{loops}$, so that different network topologies will correspond to different regions of the plane. 
The cyclomatic number $n_{loops}$  quantifies the number of independent cycles in a network, is zero for linear and branched graphs and maximum for complete graphs, where each vertex is connected with every other vertex. The Wiener index $W$ follows the opposite trend and is minimum for complete graphs, and maximum for linear ones. In particular, $W$ is able to distinguish between different degrees of branching in a tree graph. 
To compare linked components formed within system having a different numbers of rings, we normalize both indices by subtracting the minimum value and dividing by the maximum range:
\begin{equation}
    \tilde{I}=\frac{I-I_{\rm min}}{I_{\rm max} - I_{\rm min}} \,.
\end{equation}
where $I$ stand for either $W$ or $n_{loops}$. With this convention, the point $\tilde{W}=1$, $\tilde{n}_{loops}=0$ indicates a perfectly linear catenane, all points having $\tilde{n}_{loops}=0$ indicate branched catenanes, and points having $\tilde{n}_{loops}\simeq1$ would indicate fully interconnected networks, whose formation is however hindered by the channel confinement. Here we are interested in particular in the formation of branched/linear polycatenanes, circular and multi-loops polycatenanes (i.e. circular polycatenanes connected by a single edge) and Olympic networks.

The results of the analyses are reported in Fig.~\ref{fig:linked_structures} for systems with proximity-based Topo~II, $M=6$ (panels A-D) and $M=10$ rings (panels E-H), in the large channel (top row) and in the narrow channel (bottom row), for the two values of $\tau_{\rm T2}$. Different families of topologies are distinguished by different markers, with the marker size indicating the frequency with which a specific pair of indices is observed. By comparing the plots for different numbers of rings, it can be noted that, while increasing $M$ obviously results in a larger portion of the plane being explored, most topologies lie within an envelop formed by branched/linear polycatenanes (gray/blue squares), tadpoles (green and gray circles), and chorded-loop Olympic networks (orange). These plots let us further distinguish between the effect of channel size and of the Topo~II parameters on the kind of polycatenanes that are formed. Looking at the changes brought by $\tau_{\rm T2}$, the accumulation caused by faster Topo~II corresponds to an increase in the complexity of the explored topologies, with systems moving toward the bottom-right corner of the plots and more Olympic networks being present. Comparing the plots for different channel sizes one can see that going from $\Delta_r=20\sigma$ to $\Delta_r=10\sigma$ reduces the space of explored olympic network topologies, particularly for $\tau_{\rm R2}=1240\tau_{LJ}$, and, importantly, leads to the emergence of polycatenanes looping around the channel (dark red points). 
When the curvature-based criterion is included in the model, this reduces the number of olympic networks topologies observed while increasing the polycatenanes, but does not qualitatively change the other results, see SM.

\begin{figure}[h!]
    \centering
    \includegraphics[width=0.98\columnwidth]{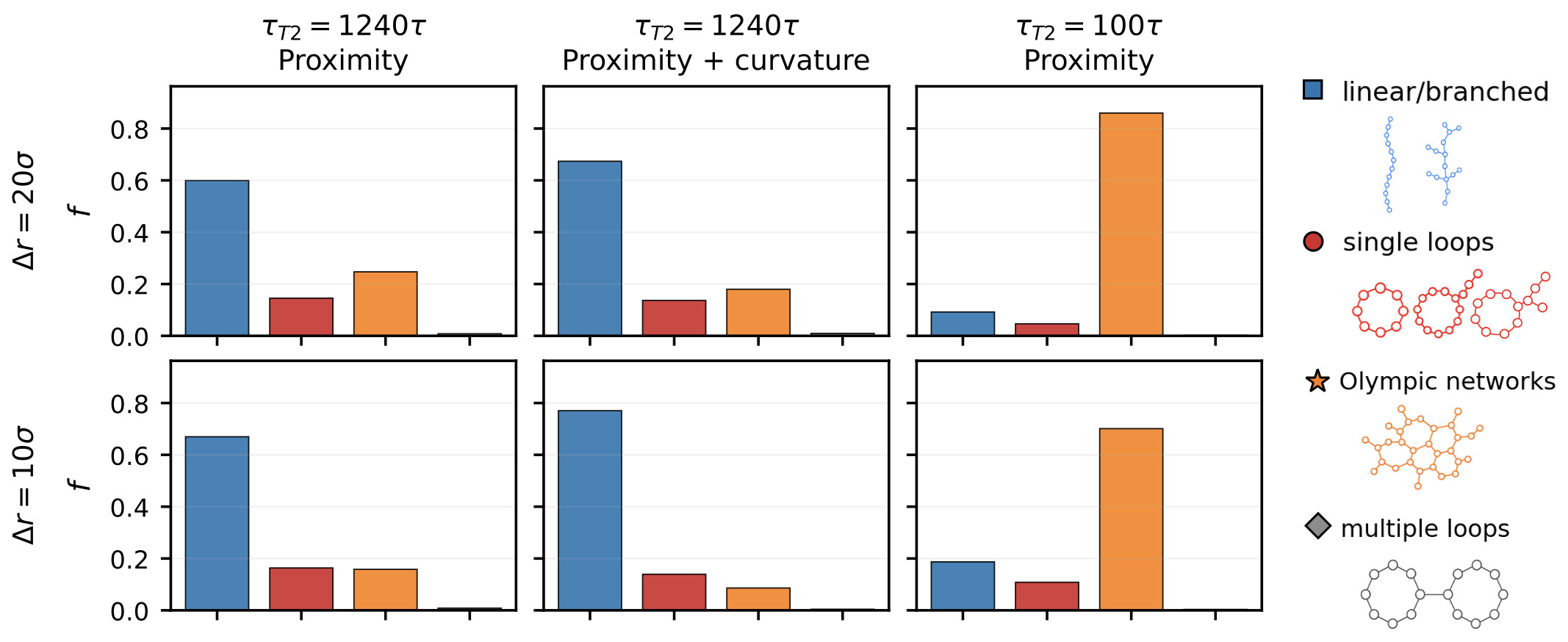}
    \caption{Histograms of explored network families for $\Delta_r=20\sigma$ (top row), $\Delta_r=10\sigma$ (bottom row), different characteristic times: $\tau_{\rm T2} = 1240\tau_{LJ}$ (left  and central column), $\tau_{\rm T2}=100\tau_{LJ}$ (right column). From left to right the columns represent linear/branched polycatenanes (blue), polycatenanes with a single loop (red), Olympic networks (yellow) and multi-loops structures (gray).}
    \label{fig:topologies_hist}
\end{figure}

The frequencies of different kinds of connected components averaged across all values of $r_m$ for the two values of $\Delta_r$ and $\tau_{\rm T2}$ are reported in Fig.~\ref{fig:topologies_hist}. Interestingly, we note that  while decreasing the action time of Topo~II significantly increases the quantity of Olympic networks, changing the channel size does not change the qualitative spectrum of topologies: linear and branched polycatenanes remain the most probable structures, followed by networks allowing a single loop. Thus, while considering a larger channel leads to a larger aggregation and unlinking number (Fig.~\ref{fig:complexity}), connected to more variability within network families, the proportion of different network types does not change significantly. In the same way, no significant change aside from a  reduction in the frequency of olympic networks and a corresponding increase in linear/branched polycatenanes is observed when including the curvature-based criterion, as shown in the two central panels of Fig.~\ref{fig:topologies_hist} (see also SM).

\section{Conclusions}
In this paper we numerically investigated  the behavior of dsDNA rings confined inside annular nanochannels  of different lengths and widths in the presence of a model Topo~II enzyme. We have shown that simple observables such as the angular span occupied by the rings and their first Fourier coefficient can be used to distinguish between self-avoiding, phantom-like and aggregated systems. We then characterized the topological properties of the catenated components explored by the systems both through the unlinking number and by mapping the catenated components to graphs. Using a complementary pair of topological indices, we mapped the catenated components to a two-dimensional space which allowed to distinguish between different graph topologies, from linear and branched polycatenanes to Olympic networks, and quantify their relative frequencies. 

Our results show that in the limit of high availability of Topo~II, the relationship between the diffusion time of the rings and the characteristic time of the Topo~II model has a strong influence on the results: if Topo~II acts faster than the rings' diffusion time, one obtains highly aggregated states composed by Olympic networks independent of the channel size. Conversely, if Topo~II is slower than the rings' diffusion time, this leads to aggregated states in a channel of size $\Delta_r=l_p$ and to phantom rings-like states in a narrower channel with $\Delta_r = l_p/2$. 
Accumulation is compatible with experimental results in system having either a very high concentration of Topo~II or some DNA condensation factor\cite{krasnow1982catenation-349,riou1985type-ii-6d0,Jeong2022, krajina2018active}. This can be explained with the fact that when rings are very close to each other there are more available sites for Topo~II to act upon, linking the rings together. Once several links are interconnected, the number of unlinking steps necessary to free one of them grows significantly, making it less probable to disentangle them. However, when Topo~II  activity is slower than the time needed for the rings to diffuse by a distance compatible with strand passage,  our results show that reducing the channel size drives the system toward an angular distribution similar to that of phantom rings. Interestingly, the possibility that DNA in presence of Topo~II could behave as a phantom chain was hypothesized by Sikorav et al.~\cite{Sikorav1994} to explain certain aspects of chromosome condensation. 

Annular channels could in principle be used to produce polycatenated materials. Our results indicate that the families of polycatenated networks explored by the system depend on the characteristic time of the Topo~II model, but do not show any significative dependence on channel size, despite the different behaviors observed for the angular distribution of  the rings. This indicates a qualitative difference between the aggregation induced by a faster Topo~II and the one related to confinement. While the first changes the kind of networks that are obtained, favoring deeply connected olympic networks over branched and linear polycatenanes, changing the degree of confinement mostly affects the size of the connected components, but not their overall topology. Interestingly, the combination of a narrow channel and a slower Topo~II, leads to the formation of circular polycatenated structure encompassing the whole channel. 

Our results hold both in the case of a proximity-only activation model and for a model that takes into account proximity plus local bending, arguably due to channel confinement forcing strand passages to happen in the bent regions of the rings, with the only relevant difference between these two models being an increased preference for linear/branched polycantenas when Topo~II preference for locally-bent regions is taken into account.

In conclusion, while the effective models investigated in this study are highly simplified, we believe that the simulated setup, if experimentally realized, would offer the possibility to characterize Topo~II catenation and decatenation behaviour, validate computational Topo~II models, and provide a way to obtain polycatenated materials with some degree of control on their topologies. 
 
\section*{Supplementary material}
In the Supplementary Material (SM) we report analytical details of the observables for the uniform distribution of points used as a reference for phantom rings, together with the extended analyses reported in Figures \ref{fig:bimodality_Fourier}, \ref{fig:complexity} and \ref{fig:linked_structures} for all cases considered. We also discussed the differences in the topological networks resulting from the simulations in which in which strand-passage events were triggered through the proximity condition alone and those with the local bending condition. 

\section*{Acknowledgments}
The authors are grateful to Davide Michieletto and An\v{z}e Bo\v{z}i\v{c} for several useful discussions and to Alex Klotz for pointing out the experimental feasibility of annular nanochannels.

\section*{Author declaration}
\subsection*{Conflict of Interest}
The authors have no conflicts to disclose.

\subsection*{Author contributions}
{\bf Carolina Palombo}: Conceptualization (equal); Formal analysis (lead); Investigation (lead); Methodology (equal); Software (lead); Writing – original draft (equal); Writing – review \& editing (equal). {\bf Davide Breoni}:  Formal analysis (supporting); Supervision(supporting); Investigation (supporting); Methodology (supporting); Writing – review \& editing (equal). {\bf Luca Tubiana}: Conceptualization (equal); Formal analysis (equal); Supervsion (lead); Investigation (supporting); Methodology (equal); Writing – original draft (equal); Writing – review \& editing (equal).

\section*{Data Availability Statement}
The data that supports the findings of this study are available from the corresponding author upon reasonable request.

\bibliographystyle{apsrev4-1}
\bibliography{references.bib}

\newpage
\onecolumngrid
\section*{Supplementary Material}
\label{par:supplements}
\onecolumngrid

\subsection*{Covered angle in the phantom-ring limit}
In the phantom-ring limit, the centers of mass of the $M$ rings are modeled as independent uniformly distributed points on a circle. Let the circumference be normalized to unity, and let $g_i$ denote the angular gaps between consecutive ordered points, with $\sum_{i=1}^M g_i = 1$. The smallest arc $S_M$ containing all points is the complement of the largest empty gap,
\[
S_M = 1 - g_{\max},
\qquad
g_{\max}=\max_i g_i \,.
\]
Following Stevens' result~\cite{Stevens1939}, the probability that all gaps are smaller than a given value $x$ is
\begin{equation}
\label{eq:Stevens}
    \mathbb{P}(g_{\max} \le x) = \sum_{k=0}^{M}(-1)^k \binom{M}{k} (1-kx)_+^{M-1},
\end{equation}
where $(y)_+=\max(y,0)$. The mean largest gap is therefore
\[
\langle g_{\max} \rangle = \int_0^1 \mathbb{P}(g_{\max}>x)\,dx=
\int_0^1 \left[ 1-\mathbb{P}(g_{\max}\le x)\right]dx \,.
\]
Substituting this expression in Eq. \ref{eq:Stevens} gives
\[
\langle g_{\max} \rangle = \sum_{k=1}^{M}(-1)^{k+1} \binom{M}{k} \int_0^{1/k} (1-kx)^{M-1}\,dx \,.
\]
Since
\[
\int_0^{1/k} (1-kx)^{M-1}\,dx = \frac{1}{kM},
\]
we obtain
\begin{equation}
    \langle g_{\max} \rangle = \frac{1}{M}\sum_{k=1}^{M}(-1)^{k+1} \binom{M}{k}\frac{1}{k} = \frac{H_M}{M}\,,
\end{equation}
where $H_M=\sum_{k=1}^{M}1/k$ is the $M$-th harmonic number. Therefore, the expected normalized covered angle is
\begin{equation}
\label{eq:covered_angle}
\left\langle \frac{S_M}{2\pi} \right\rangle
= 1 - \langle g_{\max} \rangle
= 1 - \frac{H_M}{M},
\end{equation}
Here, $g_{\max}$ denotes the largest normalized gap between consecutive points, so that the result is independent of the absolute size of the circle.


\newpage
\subsection*{Dynamics of all replicas}
\begin{figure}[h!]
    \centering
    \includegraphics[width=0.55\textwidth]{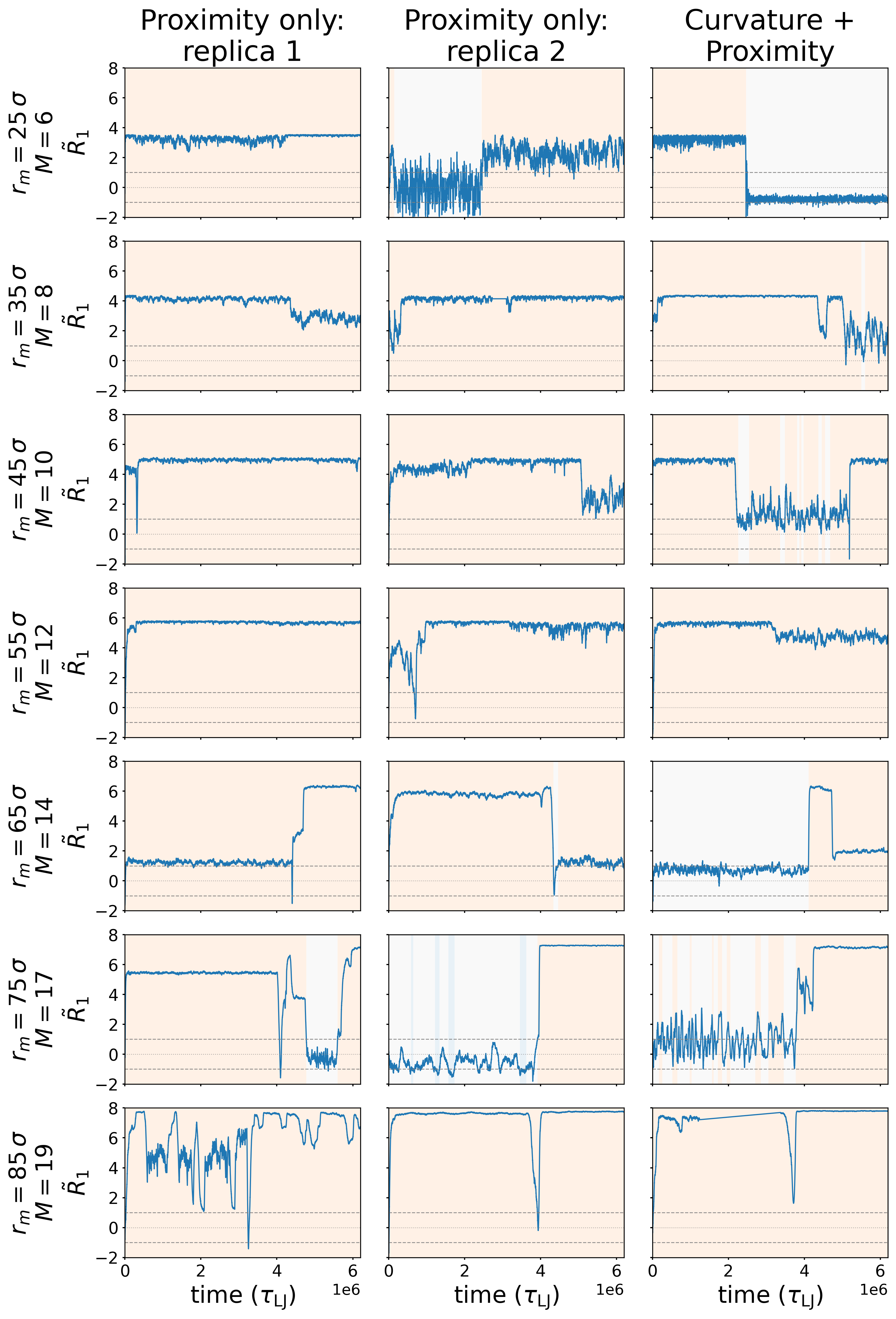}
    \hfill
    \includegraphics[width=0.4\textwidth]{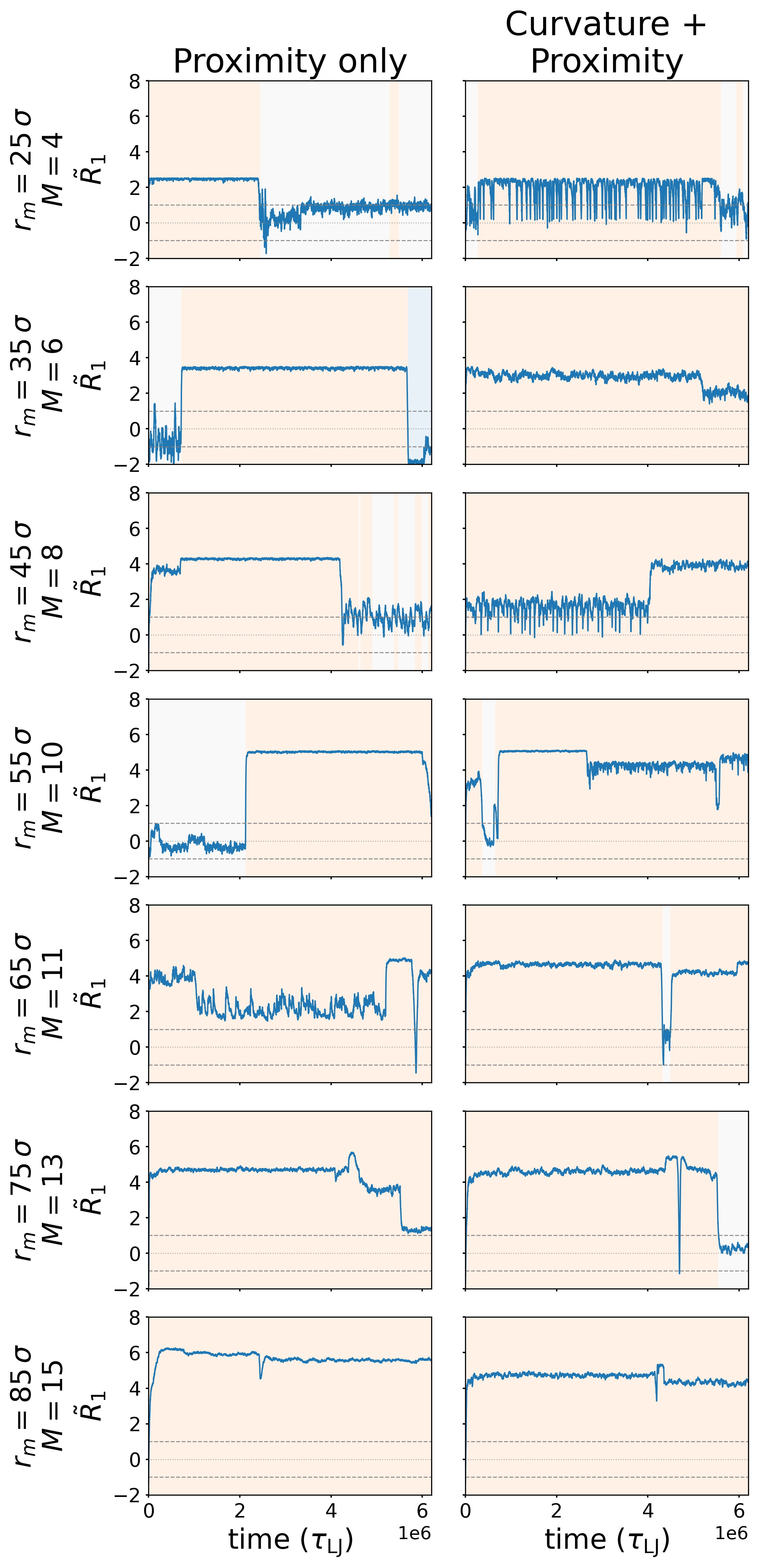}
    \caption{Time evolution of the normalized order parameter $\tilde{R}_1$ for all independent replicas and confinement geometries considered. Results are shown for channel thicknesses $\Delta r = 20\sigma$ (left) and $\Delta r = 10\sigma$ (right), for $\tau_{T2}=100\tau_{LJ}$.}
    \label{fig:trimodality_comparison_100}
\end{figure}

\begin{figure}[h!]
    \centering
    \includegraphics[width=0.49\textwidth]{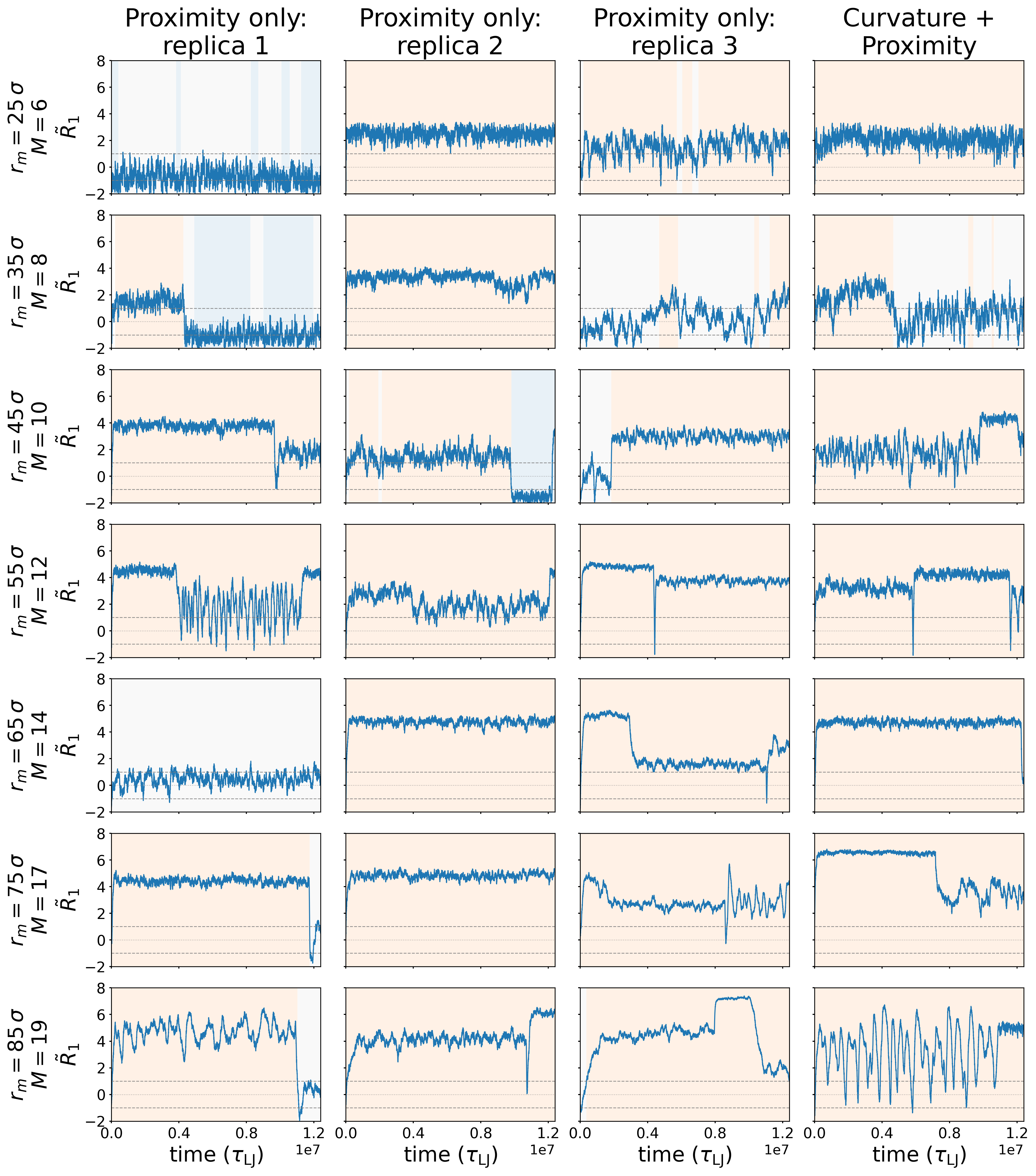}
    \hfill
    \includegraphics[width=0.49\textwidth]{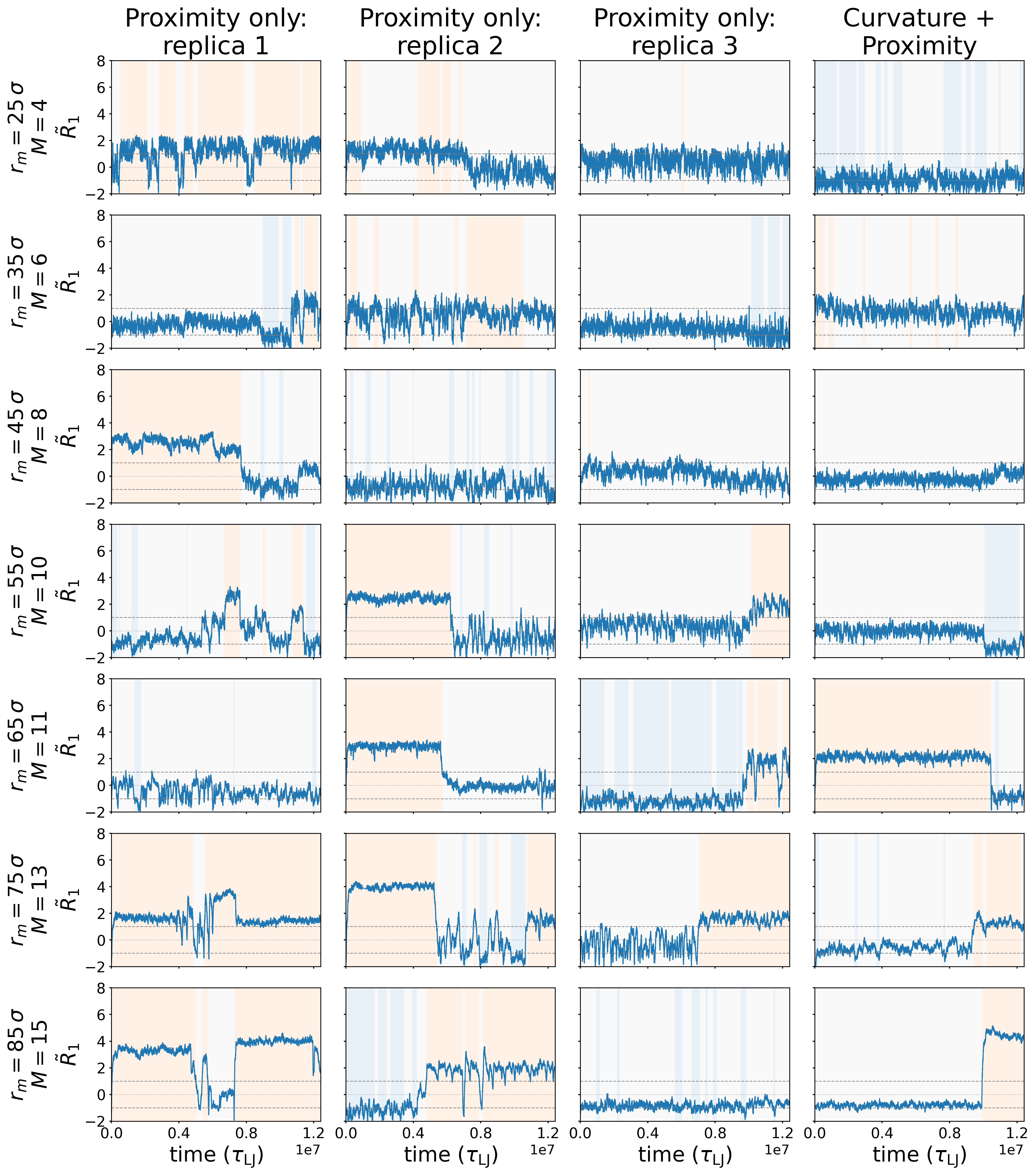}
    \caption{Time evolution of the normalized order parameter $\tilde{R}_1$ for all independent replicas and confinement geometries considered. Results are shown for channel thicknesses $\Delta r = 20\sigma$ (left) and $\Delta r = 10\sigma$ (right), for $\tau_{T2}=1240\tau_{LJ}$.}
    \label{fig:trimodality_comparison_2000}
\end{figure}

Fig. \ref{fig:trimodality_comparison_100} and \ref{fig:trimodality_comparison_2000} show the different dynamical pathways through which rings' collective distribution evolves throughout the simulations for the complete set of cases considered. As expected, for $\tau_{T2}=100\tau_{LJ}$ we overall observe a more stable accumulation state. 

\newpage
\subsection*{Relation between topological complexity and accumulation}
\begin{figure}[h!]
\centering
\includegraphics[width=0.95\textwidth]{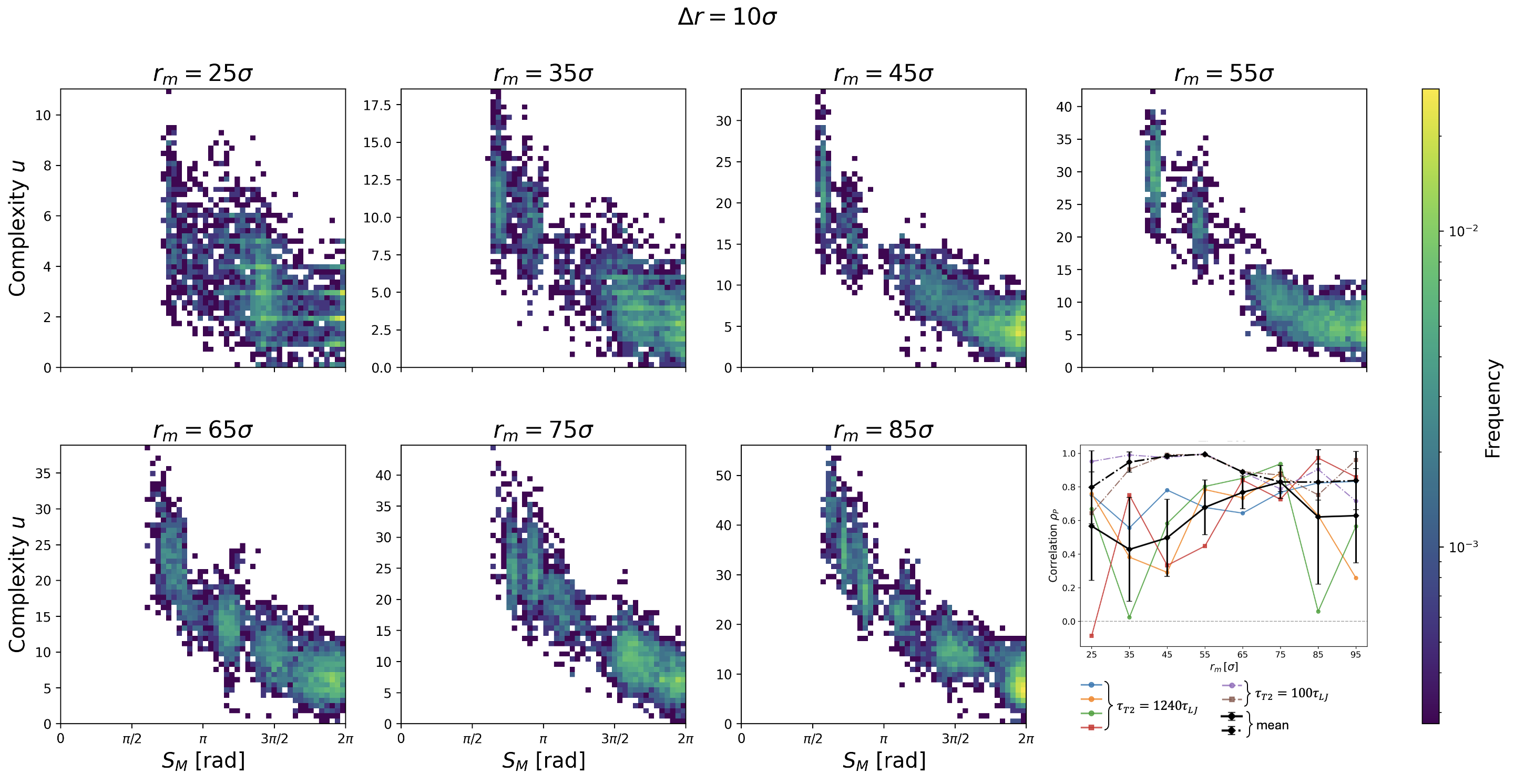}

\vspace{0.4cm}

\includegraphics[width=0.95\textwidth]{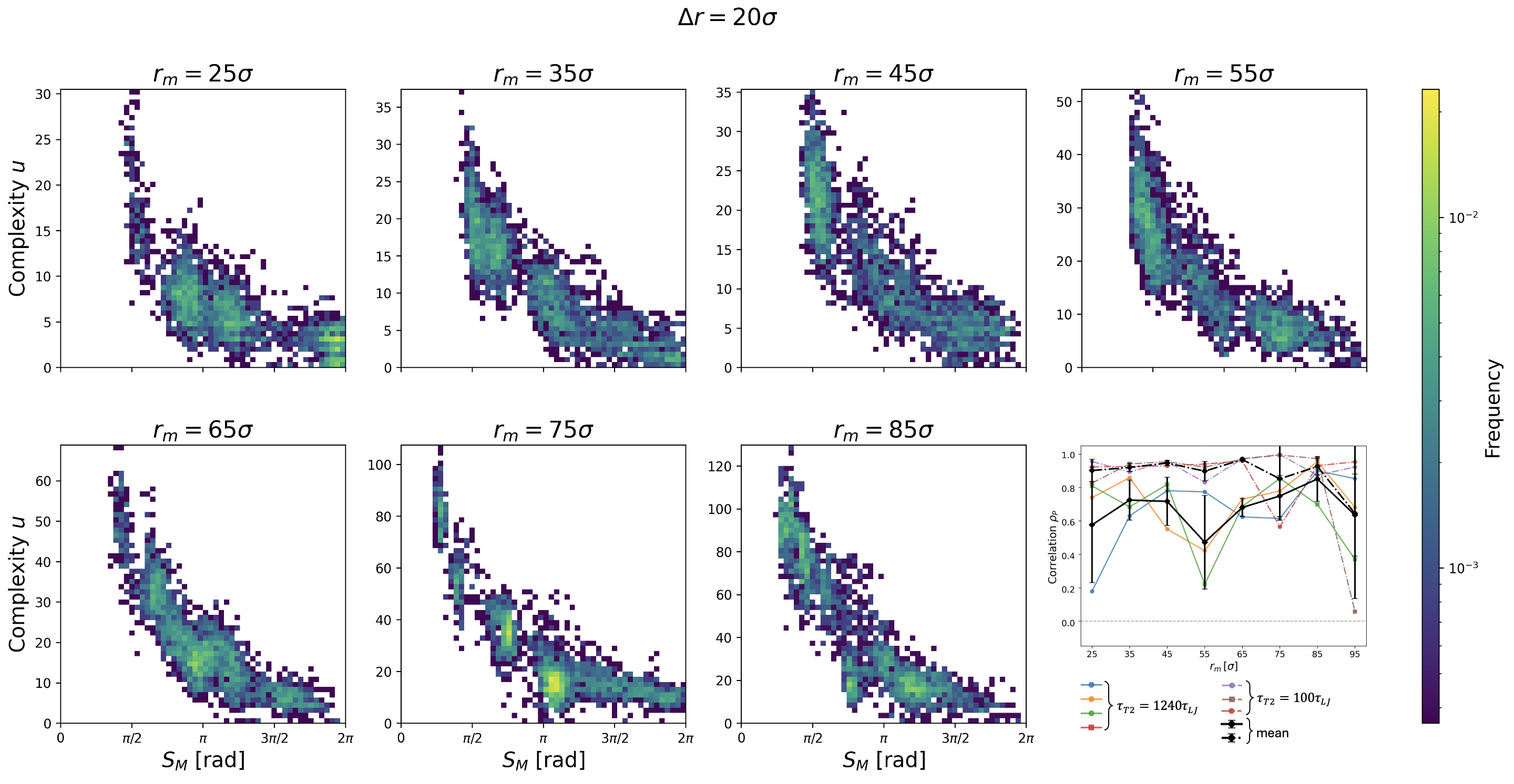}
\caption{Joint distributions of the topological complexity $u$ and the angular span $S_M$ occupied by the rings, for channel thicknesses $h=10\sigma$ (top) and $h=20\sigma$ (bottom). Each panel corresponds to a different mean channel radius $r_m$ and combines simulations performed at both values of $\tau_{T2}$. The color scale represents the normalized frequency. Across all confinement geometries, configurations with larger $u$ are preferentially associated with smaller $S_M$, highlighting the anticorrelation between network complexity and the spatial extent of the rings along the channel.}

\label{fig:correlation}
\end{figure}

\begin{figure}[h!]
\centering
\includegraphics[width=0.72\textwidth]{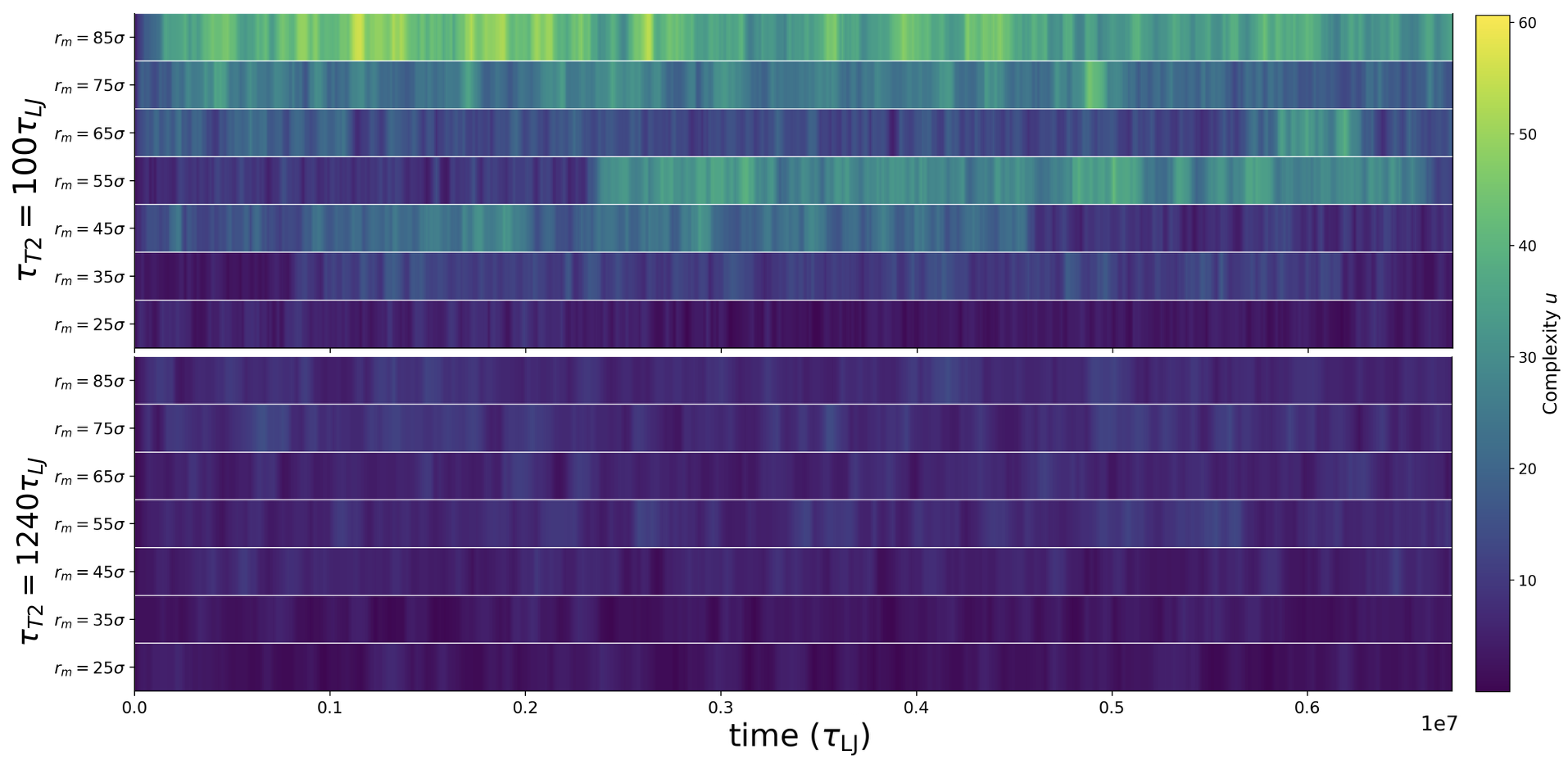}
\caption{Time evolution of the topological complexity $u$ for the different mean channel radii $r_m$, comparing simulations with $\tau_{T2}=100\tau_{LJ}$ (top) and $\tau_{T2}=1240\tau_{LJ}$ (bottom).}
\label{fig:comparison_complexity}
\end{figure}

Here, we consider the joint distribution of the network complexity $u$ and the angular span $S_M$ occupied by the rings along the channel. For each confinement volume, data from all simulations performed at both values of $\tau_{T2}$ are combined. As shown in Fig.~\ref{fig:correlation}, the distributions reveal a clear anticorrelation in all cases: configurations with larger topological complexity occupy a smaller angular fraction of the channel, whereas more spatially extended configurations are associated with lower complexity. This relation can be quantified by considering the Pearson correlation coefficient between the topological complexity $u$ and the degree of accumulation $A$, as shown in the last panel of each plot. The fact that topological complexity plays a central role in the accumulation of rings within a sector of the channel is also evident when varying the rate of Topo~II activity. By decreasing $\tau_{T2}$ from $1240\tau_{LJ}$ to $100\tau_{LJ}$, while leaving all other geometrical and model parameters unchanged, we observe a marked increase in ring accumulation together with a systematic increase in the topological complexity of the resulting ring network for all cases (Fig. \ref{fig:comparison_complexity}). 

\begin{figure}
    \centering
    \includegraphics[width=0.75\linewidth]{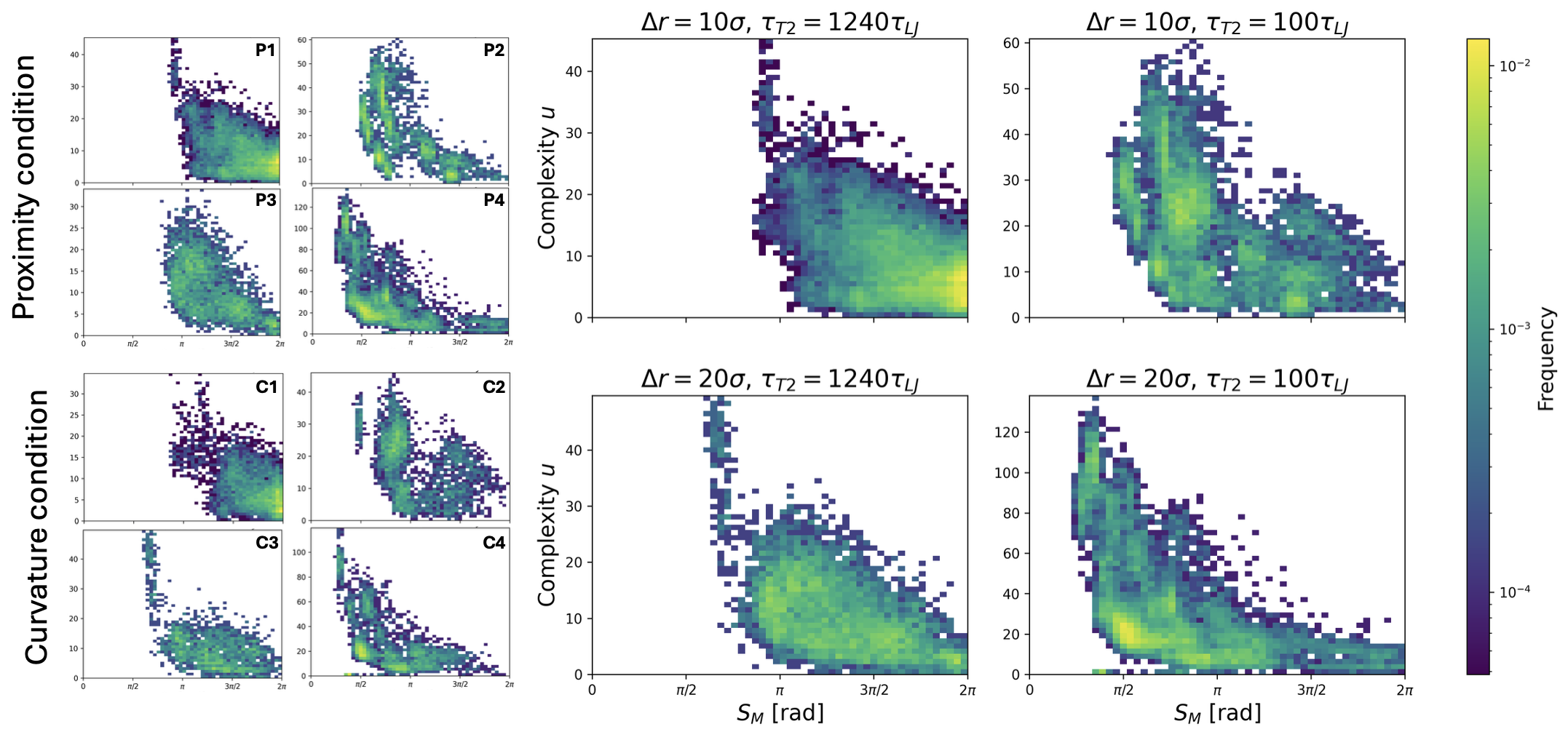}
    \caption{Joint distributions of covered angle $S_M$ and topological complexity $u$, comparing different Topo~II models to identify eligible sites. Top left: distributions from all simulations using the proximity condition alone. Bottom left: distributions from all simulations including the additional curvature condition. Right: overall distributions obtained by combining all simulations. In each case we considered $\Delta r = 10\sigma$ in panels (1, 2) and $\Delta r = 20\sigma$ in (3, 4), whereas $\tau_{T2}=1240\tau_{LJ}$ for (1, 3) and  $\tau_{T2}=100\tau_{LJ}$ for (2, 4).}
    \label{fig:complexity_comparison_model}
\end{figure}

Finally, we verified that the inclusion of the curvature condition does not qualitatively alter the relation between topological complexity and ring accumulation. In both models, configurations with smaller covered angle $S_M$ are associated with larger unlinking number $u$, and the sampled states occupy a similar region of the ($S_M,u$) plane. The main effect of the curvature criterion is a moderate reduction of the highest complexity values reached (Fig. \ref{fig:complexity_comparison_model}), consistent with its greater selectivity in identifying eligible strand-passage events.

\newpage 


\newpage
\subsection*{Topological networks for all replicas}
\begin{figure}[h!]
    \centering
    \includegraphics[width=0.9\linewidth]{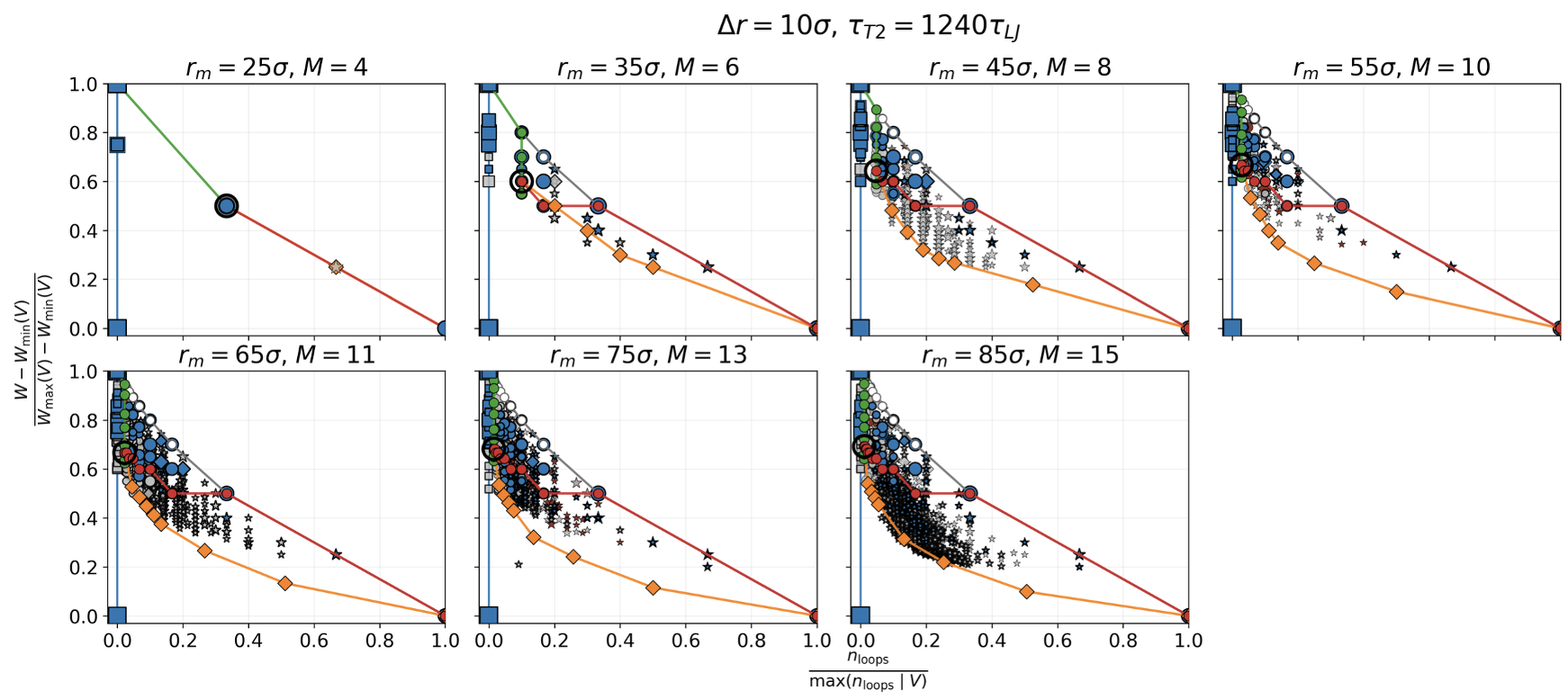}
    \caption{Comparison between the topological networks identified across the channel with $\Delta r=10\sigma$ between a replica with vicinity condition only and one in which we also consider the curvature condition, for $\tau_{T2}=1240\tau_{LJ}$. Gray markers indicates topologies observed only with the proximity condition, blue markers topologies observed both with the proximity-only condition and with the proximity-plus-curvature condition.}
    \label{fig:comparison_network_t10_2000}
\end{figure}

\begin{figure}[h!]
    \centering
    \includegraphics[width=0.9\linewidth]{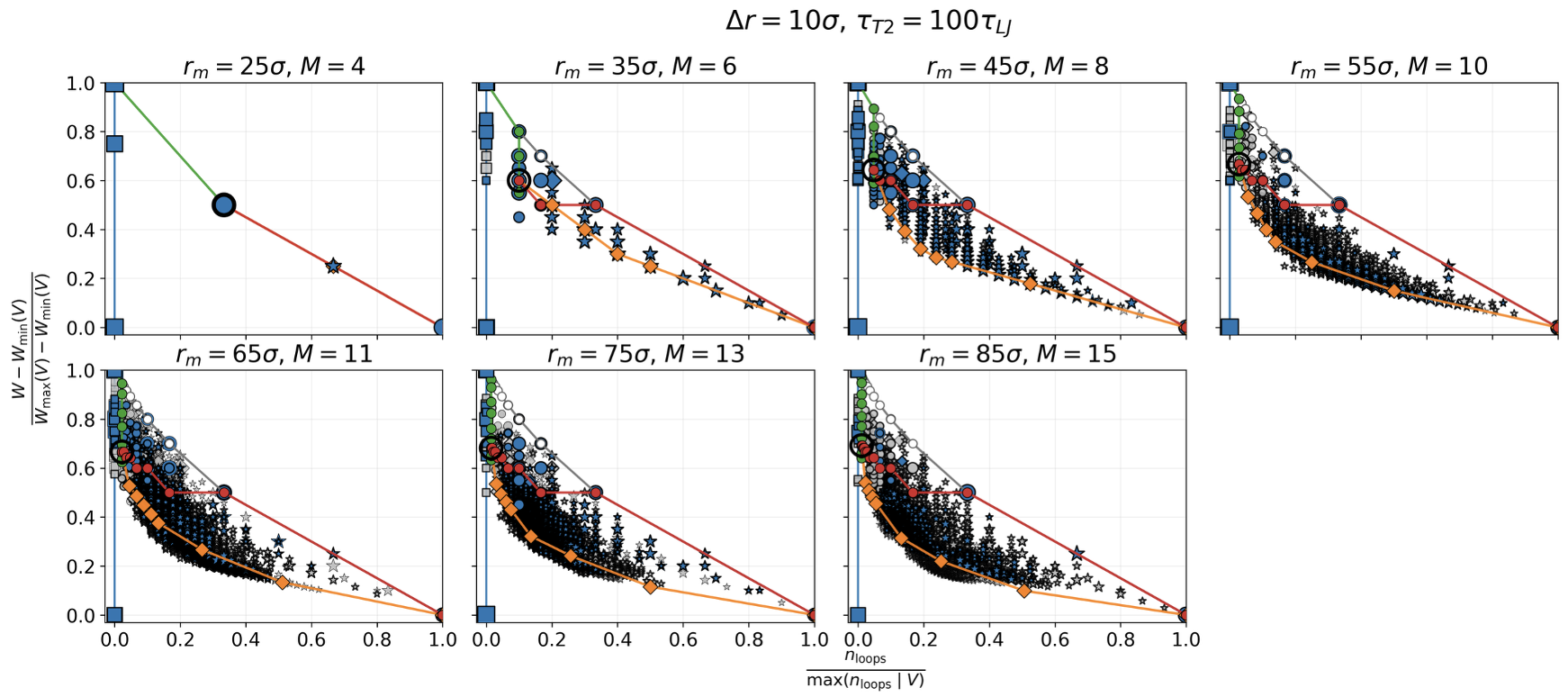}
    \caption{Comparison between the topological networks identified across the channel with $\Delta r=10\sigma$ between a replica with vicinity condition only  and one in which we also consider the curvature condition, for $\tau_{T2}=100\tau_{LJ}$. Gray markers indicates topologies observed only with the proximity condition, blue markers topologies observed both with the proximity-only condition and with the proximity-plus-curvature condition.}
    \label{fig:comparison_network_t10_100}
\end{figure}

\begin{figure}[h!]
    \centering
    \includegraphics[width=0.9\linewidth]{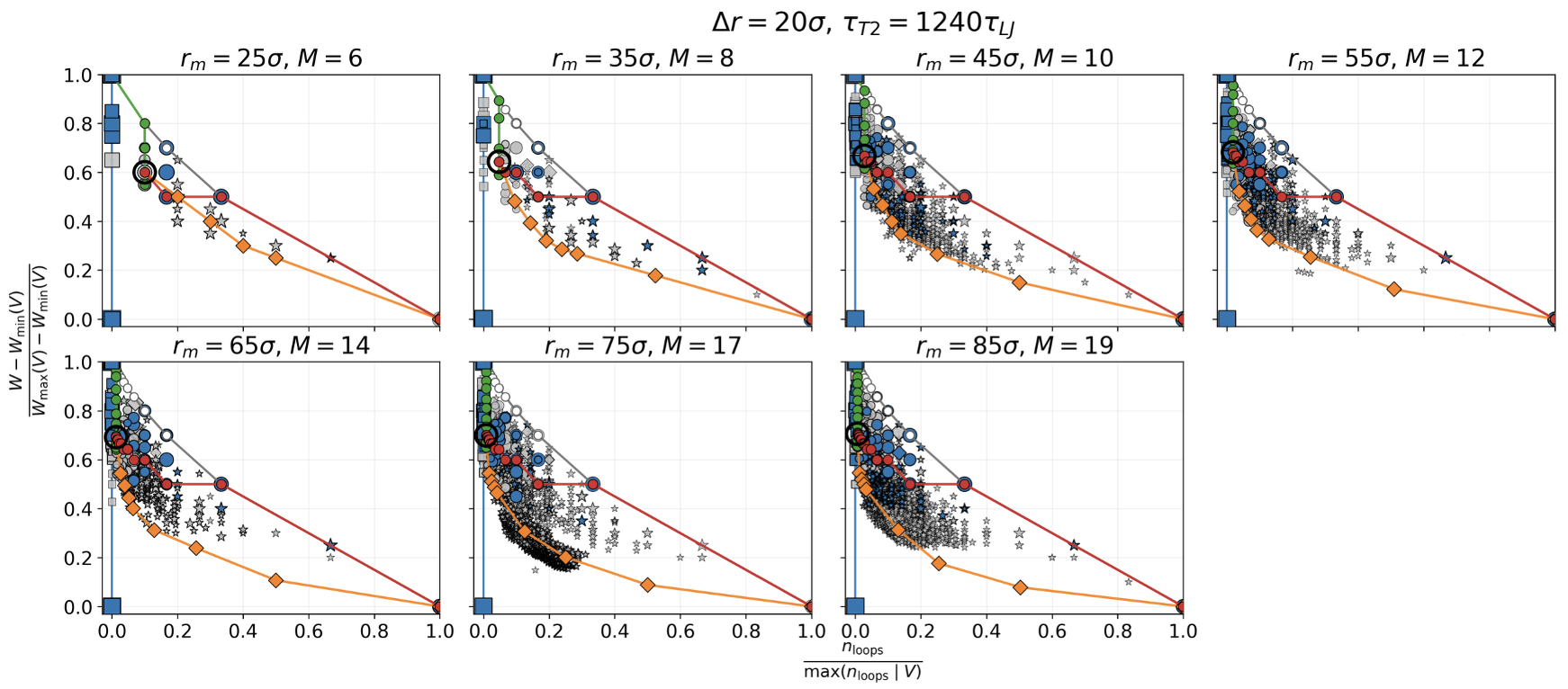}
    \caption{Comparison between the topological networks identified across the channel with $\Delta r=20\sigma$ between a replica with vicinity condition only  and one in which we also consider the curvature condition, for $\tau_{T2}=1240\tau_{LJ}$. Gray markers indicates topologies observed only with the proximity condition, blue markers topologies observed both with the proximity-only condition and with the proximity-plus-curvature condition.}
    \label{fig:comparison_network_t20_2000}
\end{figure}

\begin{figure}[h!]
    \centering
    \includegraphics[width=0.9\linewidth]{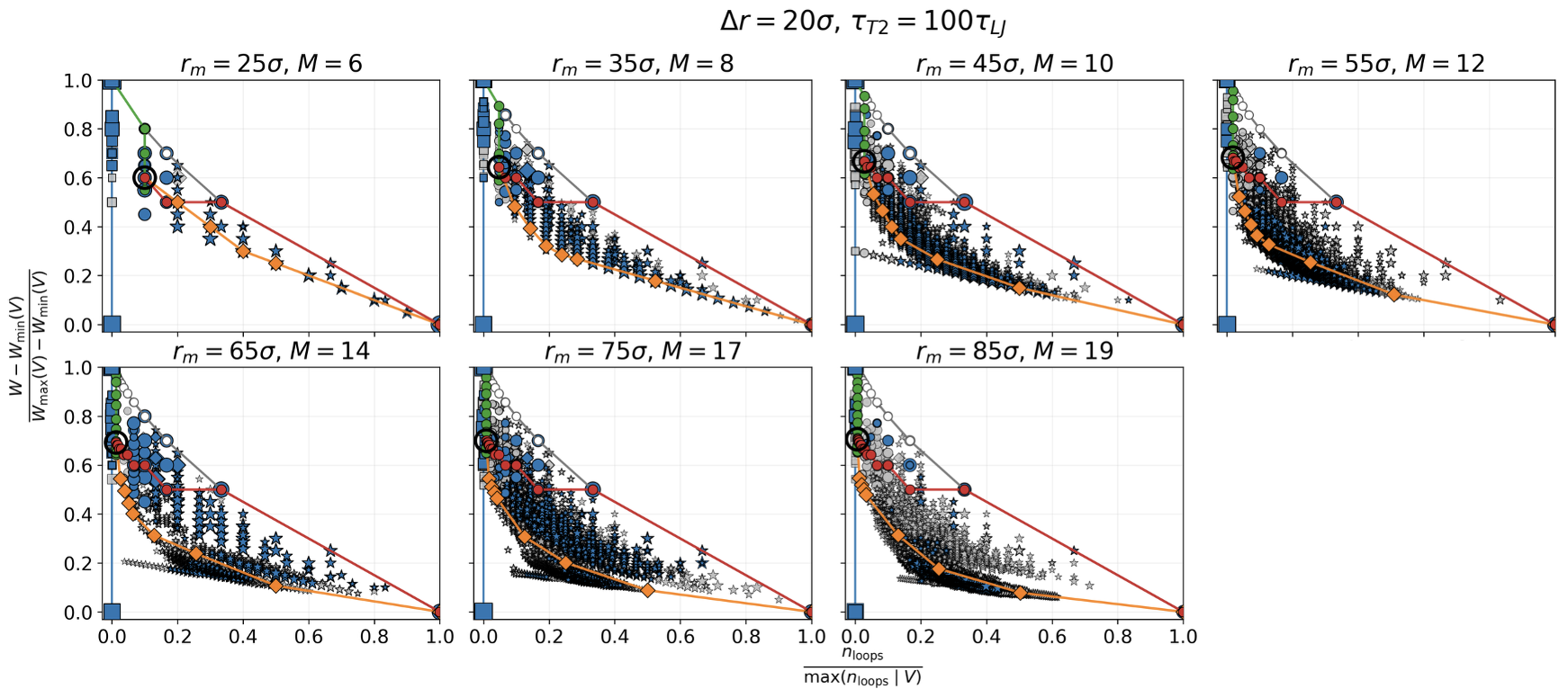}
    \caption{Comparison between the topological networks identified across the channel with $\Delta r=20\sigma$ between a replica with vicinity condition only  and one in which we also consider the curvature condition, for $\tau_{T2}=100\tau_{LJ}$. Gray markers indicates topologies observed only with the proximity condition, blue markers topologies observed both with the proximity-only condition and with the proximity-plus-curvature condition.}
    \label{fig:comparison_network_t20_100}
\end{figure}

To assess whether the additional curvature requirement modifies the types of topological structures sampled by the system, we compared the topological networks resulting from a simulation in which strand-passage events were selected using the proximity condition alone with the simulation in which the local-curvature condition was also implemented. Fig. \ref{fig:comparison_network_t10_2000}, \ref{fig:comparison_network_t10_100}, \ref{fig:comparison_network_t20_2000} and \ref{fig:comparison_network_t20_100} show the result for all confinement volumes and time $\tau_{T2}$ considered. The blue markers corresponds to the common structure between the two models compared. The main differences are observed for $\Delta r = 20\sigma$ and $\tau_{T2}=1240\tau_{LJ}$ (Fig. \ref{fig:comparison_network_t20_2000}). This observation may be a consequence of the weaker confinement, under which strongly bent configurations occur less frequently. As a result, the additional curvature criterion identifies fewer eligible strand-passage sites and therefore the events' rate is smaller than the proximity condition alone. However, for $\tau_{T2}=100\tau_{LJ}$, the ``faster'' Topo~II drives the components together more rapidly, promoting compact and interpenetrating configurations in which locally bent strand segments become more frequent. This partially compensates the additional selectivity introduced by the curvature condition and drives the system toward more complex topological configurations, yielding a distribution more similar to that obtained without the curvature criterion.

\begin{figure}[h!]
    \centering
    \includegraphics[width=0.75\linewidth]{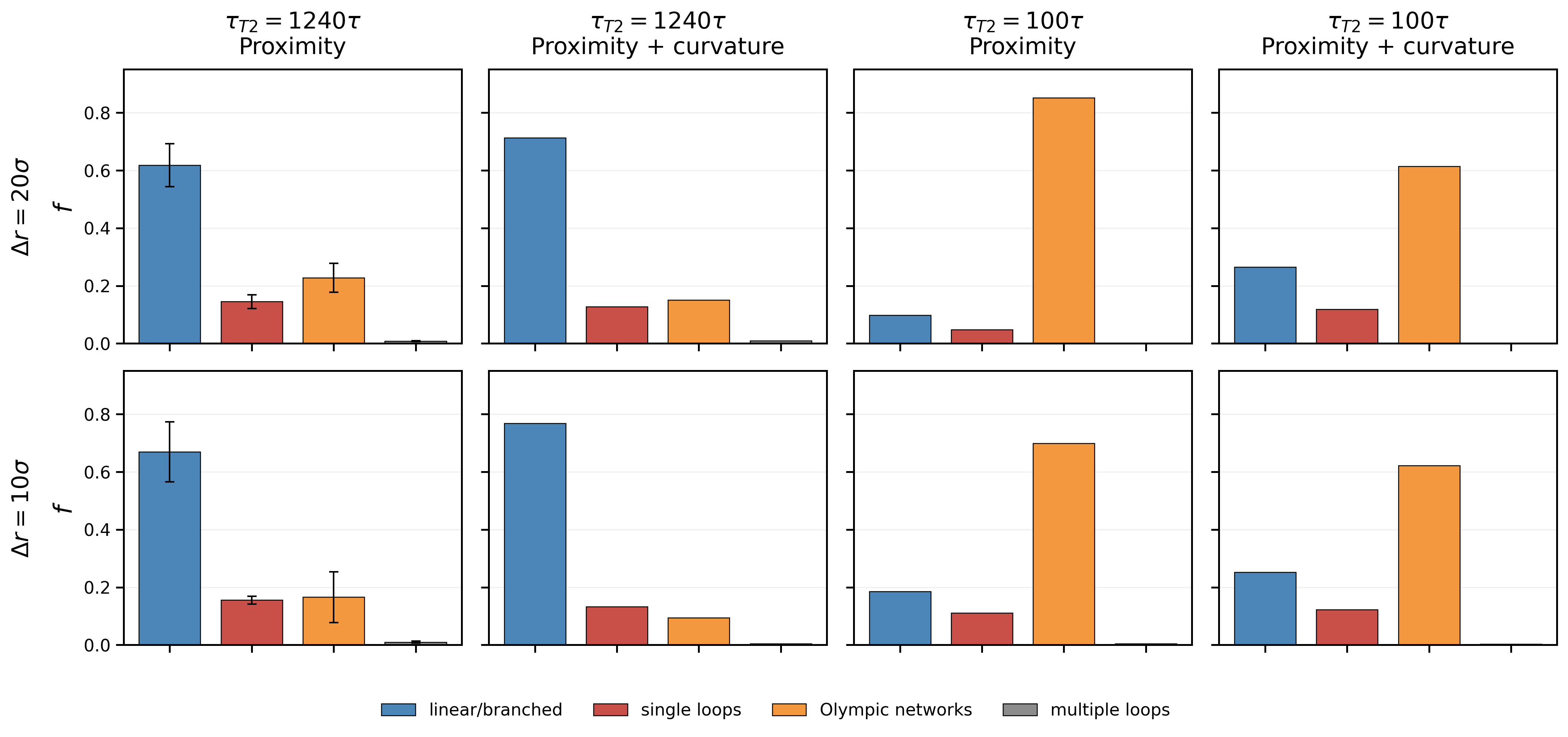}
    \caption{Histogram of the topological structure classes observed when strand-passage events are selected using either the proximity condition alone or the additional curvature criterion, for different value of the channel width $\Delta r$ and strand-passage timescales $\tau_{T2}$. Bars show the mean fraction of each class, and error bars indicate the standard deviation across replicas.}
    \label{fig:histo_comparison_all}
\end{figure}

\newpage
Nevertheless, the histograms in Fig. \ref{fig:histo_comparison_all} show that the relative occurrence of linear, single-loop, multiple-loop, and Olympic-gel structures remains qualitatively similar, indicating that the curvature criterion primarily affects the degree of topological complexity explored rather than the classes of networks formed.

\begin{figure}[h!]
    \centering
    \includegraphics[width=0.78\linewidth]{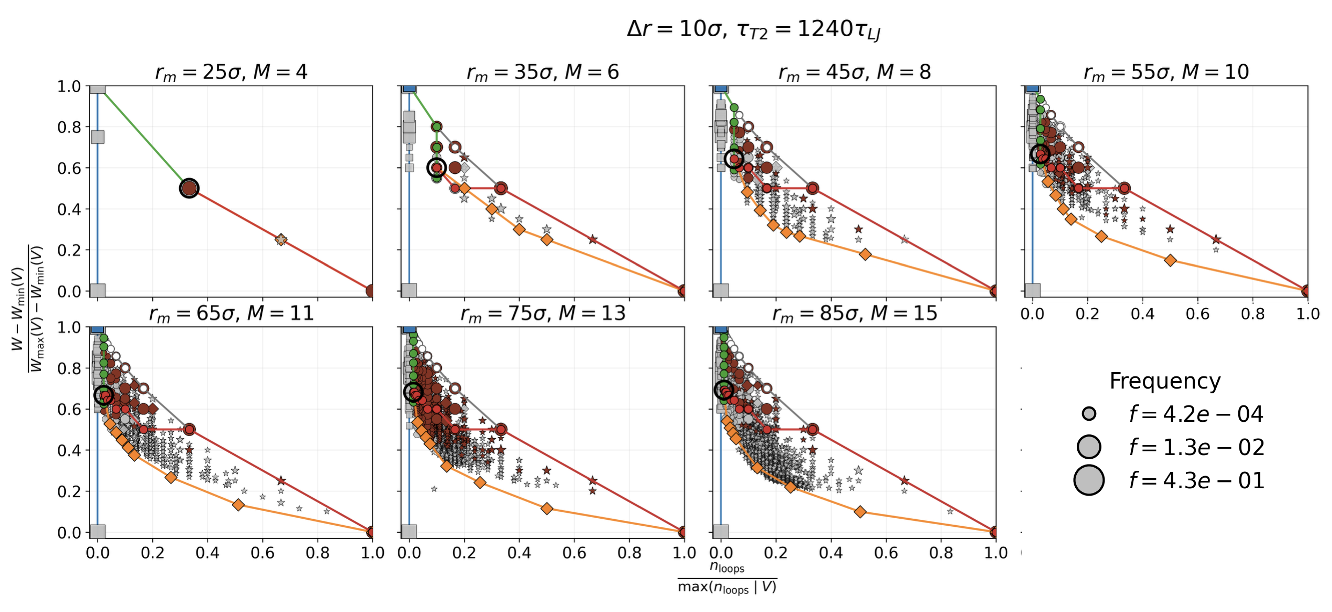}
    \caption{Topological structures observed within the channel with $\Delta r = 10\sigma$ for $\tau_{T2}=2000\tau_{LJ}$.}
    \label{fig:t10_topology_2000}
\end{figure}

\begin{figure}[h!]
    \centering
    \includegraphics[width=0.78\linewidth]{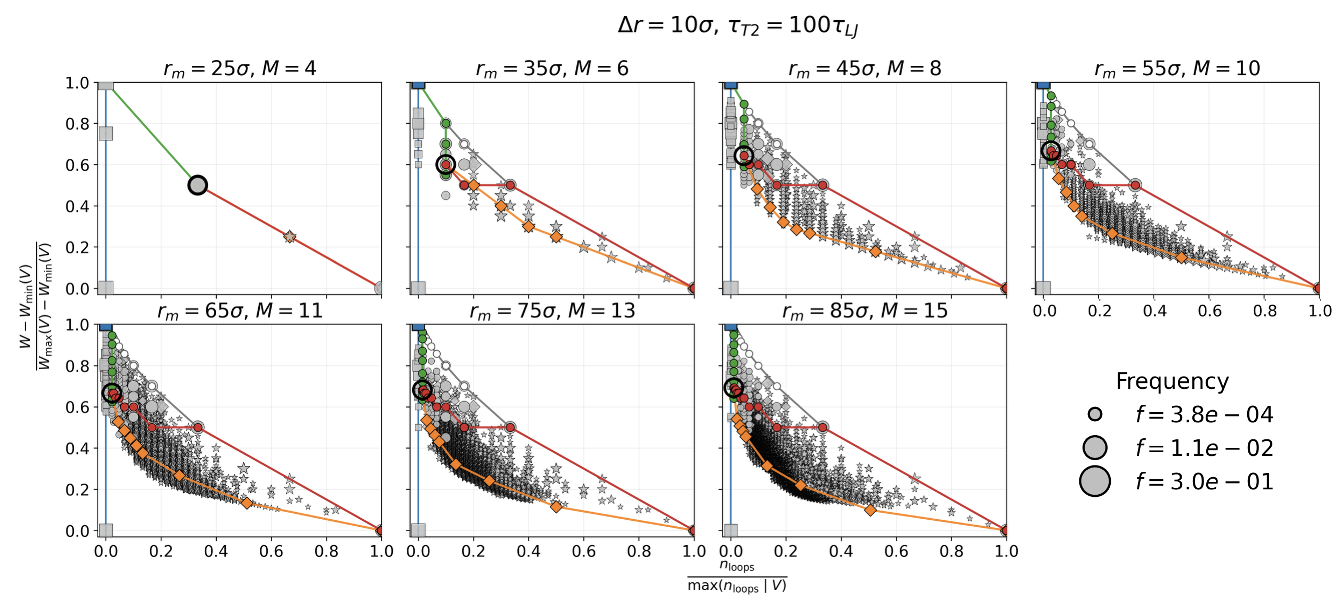}
    \caption{Topological structures observed within the channel with $\Delta r = 10\sigma$ for $\tau_{T2}=100\tau_{LJ}$.}
    \label{fig:t10_topology_100}
\end{figure}

\begin{figure}[h!]
    \centering
    \includegraphics[width=0.78\linewidth]{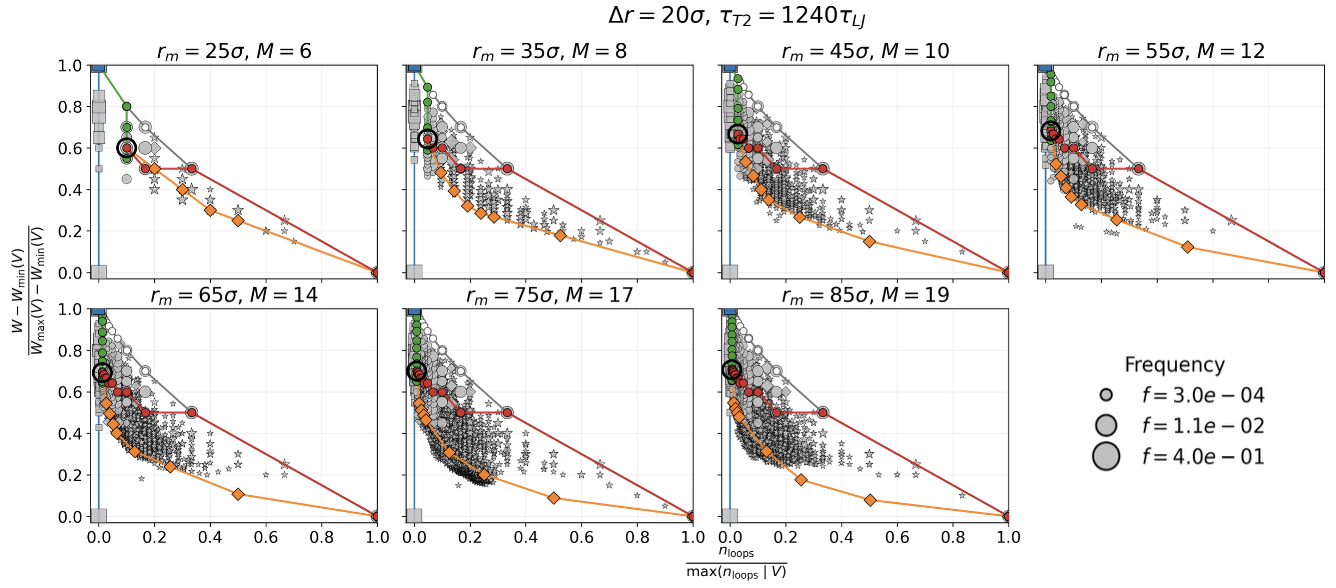}
    \caption{Topological structures observed within the channel with $\Delta r = 20\sigma$ for $\tau_{T2}=2000\tau_{LJ}$.}
    \label{fig:t20_topology_2000}
\end{figure}

\begin{figure}[h!]
    \centering
    \includegraphics[width=0.78\linewidth]{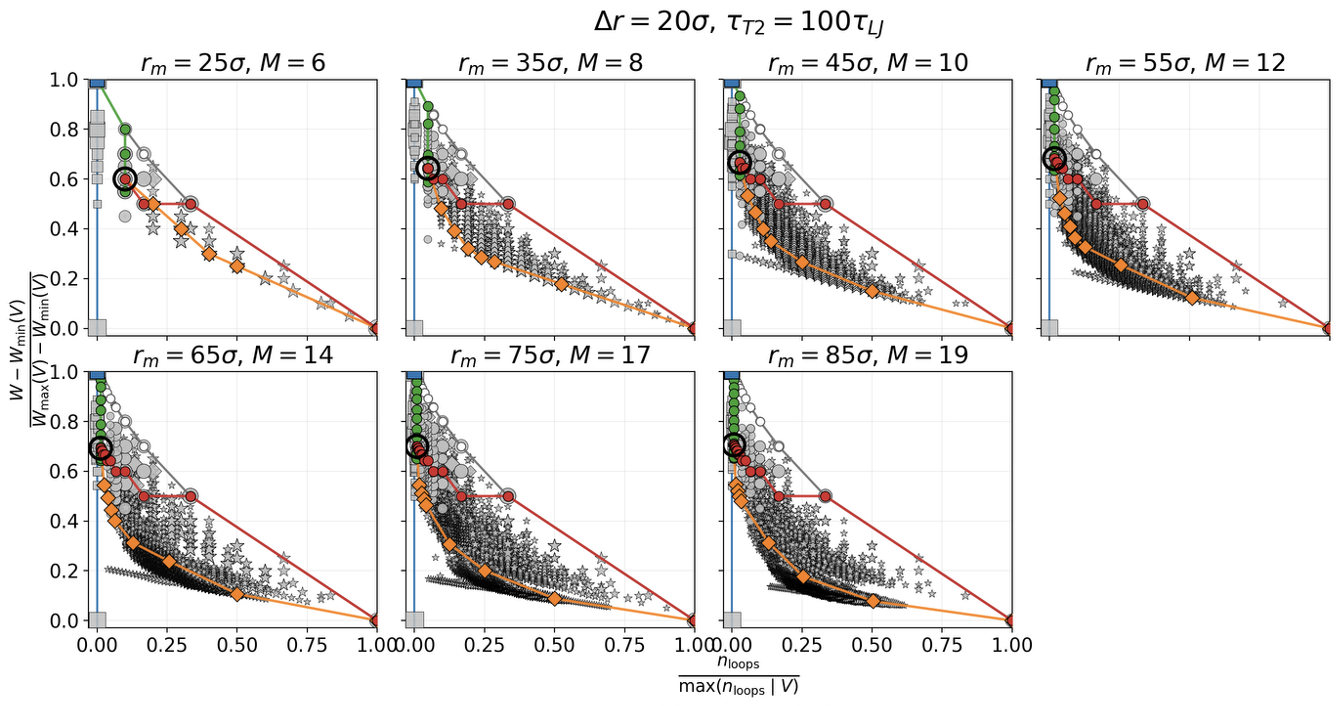}
    \caption{Topological structures observed within the channel with $\Delta r = 20\sigma$ for $\tau_{T2}=100\tau_{LJ}$.}
    \label{fig:t20_topology_100}
\end{figure}

Finally, Fig. \ref{fig:t10_topology_2000}, \ref{fig:t10_topology_100}, \ref{fig:t20_topology_2000} and \ref{fig:t20_topology_100} the set of all topological structures identified in the cases considered for different value of the channel width, $\Delta r = 10\sigma$ and $\Delta r = 20\sigma$ , and for different times $\tau_{T2}$.

\end{document}